\documentclass[a4paper,fleqn]{cas-dc}

\usepackage[numbers]{natbib}

\usepackage[utf8]{inputenc}

\usepackage[nolist,nohyperlinks]{acronym}
\usepackage[dvipsnames]{xcolor}

\usepackage{tikz}
\usetikzlibrary{fit, spy, positioning, calc, arrows, arrows.meta,
  decorations.pathreplacing, decorations.markings, decorations.text,
  shapes, tikzmark}

\usepackage{enumitem}
\usepackage{pifont}

\usepackage{amsmath,amssymb}

\usepackage{multirow}

\usepackage{listings}

\usepackage{proof, xspace}

\usepackage{fancyvrb}
\usepackage{unison}

\usepackage[justification=centering]{subfig}

\usepackage{tcolorbox}
\tcbuselibrary{listings, skins}

\usepackage{fontawesome}

\usepackage{todonotes}
\usepackage{tablefootnote}

\usepackage{calc}

\begin{acronym}
\acro{SecDivCon}{Secure Diversity by Construction}
\acro{ASLR}{Address Space Layout Randomization}
\acro{CFG}{Control Flow Graph}
\acro{BCFG}{Block Control Flow Graph}
\acro{CFG}{Control-Flow Graph}
\acro{IR}{Intermediate Representation}
\acro{MIR}{Machine Intermediate Representation}
\acro{ESV}{Extract Shared Variables}
\acro{DMG}{Datalog Model Generation}
\acro{LRA}{LLVM Register Allocation}
\acro{CP}{Constraint Programming}
\acro{COP}{Constraint Optimization Problem}
\acro{LNS}{Large Neighborhood Search}
\acro{HD}{Hamming Distance}
\acro{HW}{Hamming Weight}
\acro{MCR}{Multicompiler}
\acro{MST}{Minimum Spanning Tree}
\acro{RSA}{Rivest-Shamir-Adleman}
\acro{WCET}{Worst-Case Execution Time}
\acro{BCET}{Best-Case Execution Time}
\acro{CROG}{Constant-Resource Optimized Code Generation}
\acro{CRA}{Code-Reuse Attack}
\acro{CR}{Constant Resource}
\acro{ROT}{Register-Overwrite Transition}
\acro{MRE}{Memory-Remnant Effect}
\acro{TBL}{Transition-Based Leakage}
\acro{PSFOG}{Power Side Channel Free Optimized Code Generation}
\acro{ROP}{Return Oriented Programming}
\acro{BROP}{Blind ROP} 
\acro{JOP}{Jump Oriented Programming}
\acro{IoT}{Internet of Things}
\acro{ISA}{Instruction Set Architecture}
\acro{DPA}{Differential Power Analysis}
\acro{CPA}{Correlational Power Analysis}
\acro{PSC}{Power Side Channel}
\acro{TSC}{Timing Side Channel}
\acro{NOP}{No Operation}
\acro{SCA}{Side-Channel Attack}
\acro{CF}{Compiler Frontend}
\acro{CFI}{Control-Flow Integrity}
\acro{SA}{Security Analysis}
\acro{TA}{Timing Attacker}
\acro{PA}{Power Attacker}
\acro{RILC}{Return-into-libc}
\acro{EBB}{Empty-Block Balancing}
\acro{CBB}{Copy-Block Balancing}
\end{acronym}

\newcommand{\codebackcolor}{gray!10!white}
\newcommand{\colbackcolor}{gray!5!white}
\newcommand{\colframecolor}{white!60!black}

\definecolor{keywordCommentColor}{rgb}{0.090000, 0.55, 0.20}
\definecolor{stringColor}{rgb}{0.558215, 0.000000, 0.135316}
\definecolor{typeColor}{rgb}{0.6, 0.000000, 0.3}
\definecolor{localColor}{rgb}{0.6, 0.000000, 0.3}
\definecolor{ndkeywordColor}{rgb}{0.0, 0.558215, 0.558215}
\definecolor{commentsColor}{rgb}{0.0, 0.558215, 0.558215}
\definecolor{keywordColor}{rgb}{0.000000, 0.000000, 0.635294}
\definecolor{newgray}{rgb}{0.3, 0.3, 0.3}
\definecolor{agreen}{rgb}{0.0, 0.36, 0.15}
\definecolor{rgreen}{rgb}{0.13, 0.26, 0.12}

\newcommand{\basicCodeStyle}{\ttfamily\footnotesize\color{newgray}}

\lstdefinestyle{cstylepsc}{%
  language=C,
  frame=tb,
  numberblanklines=false,
  escapeinside=@@,
  basicstyle=\basicCodeStyle,
  framextopmargin=-10pt,
  morekeywords={u32,u8,u16},
  numbers=left,
  xleftmargin=15pt, %
  keywordstyle=\color{keywordColor}\bfseries,
  ndkeywordstyle=\color{ndkeywordColor}\bfseries,
  identifierstyle=\color{black}\ttfamily,
  commentstyle=\itshape\ttfamily\textcolor{commentsColor},
  stringstyle=\color{stringColor}\ttfamily,
  morekeywords = [2]{pub},
  keywordstyle = [2]\color{green!40!black}\itshape,
  morekeywords = [3]{key},       
  keywordstyle = [3]\color{stringColor}\itshape,
  morekeywords = [4]{mask},       
  keywordstyle = [4]\itshape,
}

\lstdefinestyle{algomine}{%
        language=python,
        basicstyle = \ttfamily\small,
        mathescape = true,
        keywordstyle=\color{keywordColor}\bfseries,
        numbers=left,
        xleftmargin=15pt, %
        identifierstyle=\color{black}\ttfamily,
        commentstyle=\itshape\ttfamily\textcolor{commentsColor},
        morekeywords = [2]{remove,insert,push,pop,top,empty,copy,extend,replace,isempty,hasSink, hasCycle, successors},
        keywordstyle = [2]\color{green!40!black}\ttfamily\bfseries,
        morekeywords = [3]{key},       
        keywordstyle = [3]\color{stringColor}\ttfamily\bfseries,
        morekeywords = [4]{mask},
        keywordstyle = [4]\color{brown!40!black}\ttfamily\bfseries,
}

\lstdefinestyle{cstyle}{%
        language=C,
        basicstyle = \ttfamily\small,
        mathescape = true,
        morekeywords = {u32,u8,u16},
        keywordstyle=\color{keywordColor}\bfseries,
        numbers=left,
        xleftmargin=15pt, %
        identifierstyle=\color{black}\ttfamily,
        commentstyle=\itshape\ttfamily\textcolor{commentsColor},
        morekeywords = [2]{pub},
        keywordstyle = [2]\color{green!40!black}\ttfamily\bfseries,
        morekeywords = [3]{key},       
        keywordstyle = [3]\color{stringColor}\ttfamily\bfseries,
        morekeywords = [4]{mask},
        keywordstyle = [4]\color{brown!40!black}\ttfamily\bfseries,
}

\lstdefinestyle{spstyle}{%
        language=C,
        basicstyle = \ttfamily\normalsize,
        mathescape = true,
        morekeywords = {u32,u8,u16},
        keywordstyle=\color{keywordColor}\bfseries,
        numbers=left,
        xleftmargin=15pt, %
        identifierstyle=\color{black}\ttfamily,
        commentstyle=\itshape\ttfamily\textcolor{commentsColor},
        morekeywords = [2]{Public},
        keywordstyle = [2]\color{green!40!black}\ttfamily\bfseries,
        morekeywords = [3]{Secret},       
        keywordstyle = [3]\color{stringColor}\ttfamily\bfseries,
        morekeywords = [4]{Random},
        keywordstyle = [4]\color{brown!40!black}\ttfamily\bfseries,
}

\lstdefinestyle{modelingstyle}{%
        basicstyle = \ttfamily\small,
        mathescape = true,
        keywords={def, pred, exists, forall, ite, return, in, sum,
        sorted, then, else, if, constraint, max, branch},
        keywordstyle=\color{keywordColor}\bfseries,
}

\lstdefinestyle{typedunistyle}{%
        basicstyle = \ttfamily\small,
        mathescape = true,
        keywords={xor, copy, in, out},
        keywordstyle=\color{keywordColor}\bfseries,
        morekeywords = [2]{pub, t0, t3},
        keywordstyle = [2]\color{green!40!black}\ttfamily\bfseries,
        morekeywords = [3]{key, t1, t4},       
        keywordstyle = [3]\color{stringColor}\ttfamily\bfseries,
        morekeywords = [4]{mask, t2, t5, t6, t7, t8, t9, t10},
        keywordstyle = [4]\color{brown!40!black}\ttfamily\bfseries,
}

\lstdefinestyle{unistyle}{%
        basicstyle = \ttfamily\small,
        mathescape = true,
        keywords={xor, copy, in, out},
        keywordstyle=\color{keywordColor}\bfseries,
        morekeywords = [2]{pub},
        keywordstyle = [2]\color{green!40!black}\ttfamily\bfseries,
        morekeywords = [3]{key},       
        keywordstyle = [3]\color{stringColor}\ttfamily\bfseries,
        morekeywords = [4]{mask},
        keywordstyle = [4]\color{brown!40!black}\ttfamily\bfseries,
}

\lstdefinestyle{armstyle}{%
        comment = [l]{@},
        escapeinside={!!},
        frame=none,
        keywordstyle=\color{keywordColor}\bfseries\ttfamily\em,
        identifierstyle=\color{black}\ttfamily,
        commentstyle=\itshape\ttfamily\textcolor{commentsColor},
        stringstyle=\color{Mahogany}\ttfamily,
        basicstyle=\ttfamily\small,
        keywords={, eor, eors, bx, str, ldr, mov, b, movs, mvns, bne, cmp,},
        morekeywords = [2]{lr,r0,r1,r2,r3,r12,sp},
        keywordstyle = [2]\color{keywordCommentColor}\bfseries, 
        numbers=left,
        xleftmargin=.08\textwidth,
        numbersep= 8pt
}

\newcommand{\toolname}{SecDivCon}

\newcommand{\powerattacker}{\ac{PA}\xspace}
\newcommand{\timingattacker}{\ac{TA}\xspace}

\newcommand{\secured}{\faLock\xspace}

\newcommand{\insecured}{\faUnlock\xspace}

\newcommand{\cmark}{\ding{51}\xspace}%

\usepackage{colortbl}
\newcolumntype{a}{>{\columncolor{gray!20!white}}r}

\newcolumntype{x}[1]{>{\centering\arraybackslash\hspace{0pt}}m{#1}}

\usepackage{hhline}

\begin{document}
\let\WriteBookmarks\relax
\def\floatpagepagefraction{1}
\def\textpagefraction{.001}

\shorttitle{Thwarting Code-Reuse and Side-Channel Attacks in Embedded Systems}

\shortauthors{Tsoupidi et al.}  

\title [mode = title]{Thwarting Code-Reuse and Side-Channel Attacks in Embedded Systems}

\author[kth] {Rodothea Myrsini Tsoupidi}
\cormark[1]
\ead{tsoupidi@kth.se}
\orcidauthor{0000-0002-8345-2752} {Rodothea Myrsini Tsoupidi}
\author[kth] {Elena Troubitsyna}
\ead{elenatro@kth.se}
\author[kth] {Panagiotis Papadimitratos}
\ead{papadim@kth.se}
\orcidauthor{0000-0002-3267-5374} {Panagiotis Papadimitratos}

\affiliation[kth]{organization={Royal Institute of Technology KTH},
            city={Stockholm},
            country={Sweden}}

\begin{abstract}
  Embedded devices are increasingly present in our everyday
  life. They often process critical information, and hence, rely on cryptographic protocols to achieve security.
  However, embedded devices remain particularly vulnerable to attackers
  seeking to hijack their operation and extract sensitive information by exploiting side channels and code reuse.
  \acfp{CRA} can steer the execution of a program to malicious
  outcomes, altering existing on-board code without direct access to
  the device memory.
  Moreover, \acfp{SCA} may reveal secret information to the attacker
  based on mere observation of the device.
  Thwarting \acp{CRA} and \acp{SCA} against embedded devices is especially challenging because embedded devices are usually resource constrained. 
  Fine-grained code diversification can hinder \acp{CRA} by
  introducing uncertainty to the binary code; while software
  mechanisms can thwart timing or power \acp{SCA}.
  The resilience to either attack may come at the price of the
  overall efficiency.
  Moreover, a unified approach that preserves these mitigations
  against both \acp{CRA} and \acp{SCA} is not available. In this paper, we  propose a novel \acf{SecDivCon} approach that tackles this challenge.
  %
  SecDivCon is a combinatorial compiler-based approach that
  combines software diversification against \acp{CRA} with software
  mitigations against \acp{SCA}.
  \ac{SecDivCon} restricts the performance overhead introduced by the
  generated code that thwarts the attacks and hence, offers a secure-by-design
  approach enabling control over the performance-security trade-off.
  Our experiments, using 16 benchmark programs, show that
  \ac{SCA}-aware diversification is effective against \acp{CRA}, while
  preserving \ac{SCA} mitigation properties at a low, controllable
  overhead.
  Given the combinatorial nature of our approach, \ac{SecDivCon} is
  suitable for small, performance-critical functions that are
  sensitive to \acp{SCA}.
  \ac{SecDivCon} may be used as a building block to
    whole-program code diversification or in a re-randomization scheme
    of cryptographic code.
    
\end{abstract}




\maketitle 

\section{Introduction}
\label{sec:intro}

Nowadays, numerous embedded devices, sensors and \ac{IoT} devices  process and
control a large variety of sensitive information.
They are typically resource-constrained and vulnerable to attacks that
aim to manipulate their operation and/or extract sensitive
information~\cite{papp_embedded_2015}.
Memory corruption vulnerabilities induce a serious security threat.
Mitigations such as data execution prevention have eradicated code
injection attacks.
Nonetheless, \acfp{CRA} achieve hijacking the control flow of a
program using a chain of executable code
snippets~\cite{shacham_geometry_2007,roemer_return-oriented_2012-2}.
These attacks target both general purpose~\cite{shacham_geometry_2007}
and embedded
devices~\cite{caballero_avrand_2016,jaloyan_return-oriented_2020,bletsch_jump-oriented_2011,salehi_microguard_2019-1}.
At the same time, the execution of embedded software may leak
information about sensitive data to the adversary via side
channels~\cite{messerges1999investigations,deogirikar2017security,devi_side-channel_2021}.
\acfp{SCA} allow an attacker to extract information from the target
device by recording side-channel information, such as execution time
or power consumption, which may depend on secret values.


Mitigating \acp{CRA} and \acp{SCA} is a double-edged challenge.
In the literature, there are solutions tailored to each of these
attacks for embedded devices.
However, there are two main drawbacks associated with combining
individual mitigations.
First, there is no guarantee that the sequential application of the
mitigations preserves the properties of each of them (see
Section~\ref{ssec:motivation}).
Second, the mitigation result may accumulate the introduced overhead
from each approach~\cite{deogirikar2017security}, which may be resource-forbidding, and thus, creates the need for overhead-aware
approaches~\cite{vu_reconciling_2021-1,tsoupidi2021constraint}.

In this paper, we address this challenge by proposing a novel approach
that combines fine-grained code diversification against \acp{CRA} with
software mitigations against \acp{SCA}.
Fine-grained code diversification~\cite{salehi_microguard_2019-1} is a
mitigation against \acp{CRA} that introduces uncertainty to the binary
code implementation, which makes the attacker payload nonfunctional.
An important advantage of fine-grained software diversification
compared to other mitigations against \acp{CRA} is its reduced
performance
overhead~\cite{pappas_smashing_2012,tsoupidi2021constraint}.
Typical mitigations against \acp{SCA} include software countermeasures
that prohibit the flow of secret information to the attacker, such as
software masking and \ac{CR} programming (see
Section~\ref{ssec:motivation}).
The compilation process may not propagate correctly these software
mitigations and, thus, the compiler needs to be aware of these
properties.

\acf{\toolname} is a combinatorial compiler-based
approach that combines code diversification against \acp{CRA} with
mitigations against \ac{TSC} and \ac{PSC} attacks.
Moreover, \ac{\toolname} uses an accurate cost model for predictable
architectures that allows control over the overall performance
overhead of the generated code.
\ac{\toolname} is appropriate for diversifying small cryptographic
core functions that may impose security threats through \acp{SCA}.
  Function-level diversification may be used for whole-program
  diversification~\cite{tsoupidi2021constraint} or in a
  re-randomization scheme~\cite{harm} against advanced code-reuse
  attacks~\cite{bittau_hacking_2014}.
%


This paper contributes with:
\begin{itemize}
\item an entirely novel (to the best of our knowledge) composable framework that combines automatic fine-grained code
  diversification and side-channel mitigations;
\item a constraint-based model to generate optimized code against
  \acp{TSC} (Section~\ref{sec:approach});
\item a secure-by-design compiler-based approach that preserves the
  properties of multiple software mitigations and enables control over
  the trade-off between performance and security
  (Section~\ref{sec:eval});
\item evidence that fine-grained automatic code diversification
  introduces side-channel leaks  (Section~\ref{ssec:rq2});
\item evidence that restraining diversity to preserve security
  measures against \acp{SCA} does not have a negative effect on
  \ac{CRA} mitigations (Section~\ref{ssec:rq4}).
\end{itemize}

\noindent
\textbf{Reproducibility:} The source code and the evaluation process
are available online:
\url{https://github.com/romits800/secdivcon_experiments}.

\section{Problem Statement and Threat Model}
\label{sec:probstatement}

Section~\ref{ssec:motivation} presents the attacks that we consider
and motivates our approach, which combines security mitigations
against \acp{CRA} and \acp{SCA}.
Section~\ref{ssec:threatmodel} presents the threat model, and finally,
Section~\ref{ssec:pstatement} defines the problem.

\subsection{Background and Motivation}
\label{ssec:motivation}

\paragraph{Code-Reuse Attacks (CRAs):}
\acp{CRA} exploit memory corruption vulnerabilities to hijack the
control flow of the victim program and take control over the
system~\cite{checkoway_return-oriented_2010,bletsch_jump-oriented_2011,riscvcfi}.
The attacker selects pieces of executable code from the victim program
memory, so-called gadgets, and stitches these gadgets together in a
chain that results in a malicious attack.
Code-reuse gadgets typically end with a control-flow instruction, such
as indirect branch, return, or call, which allows the attacker to
build a chain of gadgets.
Figure~\ref{fig:gadget-0} shows a code-reuse gadget that we extracted
using ROPGadget~\cite{ROPGadget2020} from an ARM Cortex M0 binary.
At address 0x0044, the gadget copies the value of \texttt{r2} to
register \texttt{r0} (line 1), then jumps to the next instruction
(line 2) and finally, jumps to the value of register \texttt{lr} to
the next gadget.
As demonstrated in Figure~\ref{fig:gadget-0}, code-reuse gadgets consist of
common instruction sequences that are frequently available in compiled
programs.

%
The main approaches against \acp{CRA} are \ac{CFI} and code
randomization.
\ac{CFI}~\cite{abera_c-flat_2016} enforces the dynamic execution of
the program to conform with the permitted execution paths, whereas
automatic code diversification~\cite{larsen_sok_2014} introduces
uncertainty to the location and instruction sequence of the gadgets in
the program memory.
\ac{CFI} may be impractical for small, resource-constrained devices
due to the diversity of embedded hardware and the increased
overhead~\cite{nyman_cfi_2017} in small, often battery-operated
devices.
Automatic software diversification provides an efficient mitigation
against
\acp{CRA}~\cite{larsen_sok_2014,tsoupidi2021constraint,caballero_avrand_2016}.
Figure~\ref{lst:gadgets} shows two gadgets in two diversified program
variants.
Figure~\ref{fig:gadget-0} and \ref{fig:gadget-1} illustrate that they
differ in the first instruction, which copies the content of the register
\texttt{r2}/\texttt{r3} to \texttt{r0}.
An attacker that has designed an attack that uses the first gadget at
address \texttt{0x0044} to move an attacker-controlled value from
\texttt{r2} to \texttt{r0} will fail if the victim uses the second
gadget.
There are different ways to diversify software and distribute it to
the end users.
In this paper, we consider the app store model~\cite{larsen_sok_2014},
where a centralized repository distributes precompiled code variants
to each end user.
%

\begin{figure}
  \begin{tcolorbox}[colback=\codebackcolor, colframe=\codebackcolor,
      top=-15pt, bottom = -15pt, right = 0pt, left = 0pt]
    \subfloat[Gadget 1\label{fig:gadget-0}]{
    \begin{minipage}[b]{0.48\textwidth}
\begin{lstlisting}[style=armstyle]
0x0044 : mov r0, r2
0x0046 : b #0x48
0x0048 : bx lr
\end{lstlisting} 
      \vspace{-5pt}
    \end{minipage}
    } \subfloat[Gadget 2\label{fig:gadget-1}]{
      \begin{minipage}[b]{0.48\textwidth}
\begin{lstlisting}[style=armstyle]
0x0044 : mov r0, r3
0x0046 : bx lr
0x0048 : ...
\end{lstlisting}
      \vspace{-5pt}
      \end{minipage}
    }
  \end{tcolorbox}
  \vspace{10pt} 
\caption{\label{lst:gadgets} Two diversified gadgets in ARM Thumb
  extracted from Figure~\ref{lst:cr} using ROPGadget}
\end{figure}

\begin{figure}
\centering
    \begin{tcolorbox}[colback=\codebackcolor, colframe=\codebackcolor,
      top=-8pt, bottom = -25pt, right = 0pt, left = 0pt]
      \subfloat[Original C code\label{lst:masked}]{
    \begin{minipage}[b]{0.50\textwidth}
\begin{lstlisting}[style=cstyle,numbersep=3pt,xleftmargin=.02\textwidth]
u32 Xor(u32 pub, u32 key, u32 mask) {
 u32 mk = mask ^ key;
 u32 t = pub ^ mk;
 return t;
}
\end{lstlisting}
    \vspace{-5pt}
    \end{minipage}
} \subfloat[Compiler-induced masking removal\label{lst:unxor}]{
      \begin{minipage}[b]{0.48\textwidth}
\begin{lstlisting}[style=cstyle, firstnumber=2, firstline=2,
          numbersep=3pt,
          xleftmargin=.02\textwidth]
 u32 t1 = pub ^ key;
 u32 t2 = mask ^ t1;
 return t2;
}
\end{lstlisting}
      \vspace{-5pt}
      \end{minipage}
    }
  \end{tcolorbox}
  \vspace{12pt}
\caption{\label{lst:masked_all} Masked exclusive OR implementation}
\end{figure}

%
\paragraph{Side-Channel Attacks (SCAs):}

Usually, embedded devices use cryptographic algorithms, which are vulnerable to
\acp{SCA}~\cite{kocher_timing_1996-2,messerges1999investigations,brier_correlation_2004-1}.
These attacks allow the adversary to extract information about secret
values by measuring the execution time -- called timing side channel 
(\ac{TSC})~\cite{brumley_remote_2011} or the power consumption -- called power side channel 
(\ac{PSC})~\cite{randolph_power_2020} of the target device.
For example, a publicly installed camera or a smartwatch may be
physically exposed to malicious actors that are able to measure the
power consumption of the device or the execution time of cryptographic
tasks to infer cryptographic keys and retrieve information about
sensitive data.

\paragraph{\acfp{PSC}:}
\ac{PSC} attack is a \ac{SCA} that uses the power traces of the target
device to extract secret information~\cite{xu_side-channel_2018}.
A mitigation approach to protect against \acp{PSC} is software
masking.
Consider the code in Figure~\ref{lst:masked}.
Function \texttt{Xor} applies software masking to an exclusive or
(xor) operation.
The program takes three inputs, \texttt{pub}, which is a public value,
\texttt{key}, which is a secret value, and \texttt{mask}, which is a
randomly generated value.
At line 2, the code performs an exclusive or operation between
\texttt{mask} and \texttt{key} to randomize the secret value.
At line 3, the implementation performs an additional exclusive or
operation between the previous result and value \texttt{pub}.
Figure~\ref{lst:leak} shows two machine implementations of the code in
Figure~\ref{lst:masked} in ARM Thumb.
The first implementation in Figure~\ref{fig:xor-seccon-0} performs the
first xor operation at line 2 and stores the result in register
\texttt{r1} and the then, performs the second xor operation at line 3
and stores the result in register \texttt{r0}.
The second implementation in Figure~\ref{fig:xor-seccon-1} is identical
to the first one, apart from the first xor operation at line 2, where
the result is copied to register \texttt{r2}.
The power leakage of the program depends on register-value
transitions, \ac{ROT}~\cite{papagiannopoulos_mind_2017}, based on the
\ac{HD} model~\cite{brier_correlation_2004-1}, which is widely used
for designing \ac{PSC} attacks and
defenses~\cite{brier_correlation_2004-1,papagiannopoulos_mind_2017}.
The leakage using the \ac{HD} model depends on the exclusive or of the
previous value of a register and the new value.
Thus, as shown in Figure~\ref{fig:xor-seccon-0}, the leakage depends on two
transitions of registers \texttt{r0} and \texttt{r1}, with values
$\texttt{r1}_{old} \oplus \texttt{r1}_{new} = \texttt{key} \oplus
(\texttt{key} \oplus \texttt{mask}) = \texttt{mask}$ and
$\texttt{r0}_{old} \oplus \texttt{r0}_{new} = \texttt{pub} \oplus
(\texttt{key} \oplus \texttt{mask} \oplus \texttt{pub}) =
\texttt{mask}\oplus \texttt{key}$.
None of the values depends on a secret value because both values are
randomized with \texttt{mask}.
However, in Figure~\ref{fig:xor-seccon-1} we have a different leakage,
$\texttt{r2}_{old} \oplus \texttt{r2}_{new} = \texttt{mask} \oplus
(\texttt{key} \oplus \texttt{mask}) = \texttt{key}$ and
$\texttt{r0}_{old} \oplus \texttt{r0}_{new} = \texttt{pub} \oplus
(\texttt{key} \oplus \texttt{mask} \oplus \texttt{pub}) =
\texttt{mask}\oplus \texttt{key}$.
The first value leaks information about the \texttt{key}, which is
secret (highlighted in Figure~\ref{fig:xor-seccon-1}).
This means that the implementation in Figure~\ref{fig:xor-seccon-1}
leaks secret information.
Thus, the embedded devices that use this variant may be vulnerable to
\acp{PSC}.
Notice that these leaks correspond to the implementation of
  the code in Figure~\ref{lst:masked}.
  If the compiler middle-end takes advantage of the associativity
  property of the exclusive-or operation to change the order of the
  exclusive-or operations, as we see in Figure~\ref{lst:unxor}, then,
  there will be a leak regardless of the register allocation (see
  Section~\ref{ssec:ct}).

\begin{figure}
  \begin{tcolorbox}[colback=\codebackcolor, colframe=\codebackcolor,
      top=-8pt, bottom = -15pt, right = 0pt, left = 0pt]
    \subfloat[Secure\label{fig:xor-seccon-0}]{
    \begin{minipage}[b]{0.48\textwidth}
\begin{lstlisting}[style=armstyle,firstline=28,lastline=31]
@ r0: pub, r1: key, r2: mask
eors	r1, r2
eors	r0, r1
bx	lr
\end{lstlisting}
      \vspace{-5pt}
    \end{minipage}
    } \subfloat[Insecure\label{fig:xor-seccon-1}]{
      \begin{minipage}[b]{0.48\textwidth}
\begin{lstlisting}[style=armstyle,firstline=29,lastline=31,firstnumber=2]
eors	!\tikzmark{r21}!r2!\tikzmark{r22}!, r1
eors	r0, r2
bx	lr
\end{lstlisting}
      \begin{tikzpicture}[remember picture,overlay]
\draw[fill=red, rounded corners, opacity=0.3]
  ([shift={(-3pt,1.5ex)}]pic cs:r21) 
  rectangle 
  ([shift={(2pt,-0.65ex)}]pic cs:r22);
\end{tikzpicture}
      \vspace{-14pt}
    \end{minipage}
    }
  \end{tcolorbox}
  \vspace{5pt} 
\caption{\label{lst:leak} Two program variants of
  Figure~\ref{lst:masked} for ARM Cortex M0 }
\end{figure}

\paragraph{\acfp{TSC}:}
\ac{TSC} attack is another type of \ac{SCA}, where the attacker
measures the execution time during the execution of a program to infer
secret information.
For example, Figure~\ref{lst:cr} shows a simple program that contains a
timing vulnerability.
In particular, at line 3 there is a branch that compares the value of
\texttt{key} and the value of \texttt{pub}.
The attacker knows and may control the value of \texttt{pub}, whereas
\texttt{key} is a secret value.
If the result of the comparison is \texttt{true}, then the observed
execution time will be longer than when the result is \texttt{false}.
Thus, an attacker who can measure the execution time of the code and
knows the value of \texttt{pub} is able to infer information about the
value of \texttt{key}.

\begin{figure}
  \begin{tcolorbox}[colback=\codebackcolor, colframe=\codebackcolor,
      top=-8pt, bottom = -10pt, right = 0pt, left = 0pt]
\begin{lstlisting}[style = cstyle]
  u8 check_bit(u8 pub, u8 key) {
    u8 t = 0;
    if (pub == key) t = 1;
    return t;
  }
\end{lstlisting}
\end{tcolorbox}
\caption{\label{lst:cr} Program with secret-dependent branching }
\end{figure}

%
The \acf{CR} policy is a software-based mitigation approach against
\ac{TSC} attacks that aims at eliminating timing
leaks~\cite{ngo_verifying_2017-1}.
The \ac{CR} policy allows secret-dependent branches, as long as the
different execution paths require the same execution time.
The implementation of \ac{CR} code is hardware specific because the
same instruction may take a different number of cycles in different
processor implementations.
Figure~\ref{lst:leak_cr} shows two machine implementations for ARM
Cortex M0 that preserve the \ac{CR} policy of the program in
Figure~\ref{lst:cr}, where the \texttt{if} branch in Figure~\ref{lst:cr}
is balanced with an \texttt{else} branch.
In Figure~\ref{fig:pass-seccon-0}, the first basic block (lines 3-5)
initializes \texttt{t} (line 3) and compares these two input values
(lines 4-5).
If the result of the comparison is \texttt{true} (taken branch), the
execution jumps to the third branch, \texttt{.LBB0\_2}, and the branch
operation takes three cycles.
If the result of the comparison is \texttt{false} (not-taken branch),
the execution continues to the second branch (\texttt{@BB\#1}) and the
branch operation takes just one cycle.
To balance the two branches, the code generation considers the branch
overhead for taken branches and the latency of every instruction,
which is three cycles for the unconditional branch, \texttt{b}, and
one cycle for the move instruction, \texttt{mov}.
In particular $t(\texttt{@BB\#1}) + 2 = t(\texttt{.LBB0\_2})$, where
$t(b)$ is the execution time of the body of basic block $b$ and $+2$
corresponds to the branch overhead on a taken branch.
Figure~\ref{fig:pass-seccon-1} shows another machine implementation of
the code in Figure~\ref{fig:pass-seccon-0} that also preserves the same
constraint as Figure~\ref{fig:pass-seccon-0}.
The main differences in Figure~\ref{fig:pass-seccon-1} concern the
register assignment.
For example, Figure~\ref{fig:pass-seccon-1} introduces additional
\texttt{mov} instructions (lines 8, 11, and 12) to transfer values
from one hardware register to another.
Without the constraint that enforces the equality of execution time
for the two branches, a randomization procedure, may break the \ac{CR}
policy, for example, by adding one \ac{NOP} instruction in
\texttt{.LBB0\_2}.

\begin{figure}
  \begin{tcolorbox}[colback=\codebackcolor, colframe=\codebackcolor,
      top=-8pt, bottom = -20pt, right = 0pt, left = 0pt]
    \subfloat[Secure variant 1\label{fig:pass-seccon-0}]{
    \begin{minipage}[b]{0.48\textwidth}
\begin{lstlisting}[style=armstyle,firstline=27,lastline=41]
@ r0: pub, r1: key
@ BB#0:                 
  movs  r2, #0  
  cmp   r1, r0  
  bne   .LBB0_2 
@ BB#1:                 
  movs  r0, #1  
  b     .LBB0_3 
.LBB0_2:                
  mov   r0, r2  
  movs  r1, #1  
.LBB0_3:                
  bx    lr
...
...
\end{lstlisting}
    \end{minipage}
    } \subfloat[Secure variant 2\label{fig:pass-seccon-1}]{
      \begin{minipage}[b]{0.48\textwidth}
\begin{lstlisting}[style=armstyle,firstline=28,lastline=41,firstnumber=2]
@ BB#0:                                                                                
  movs  r2, #0                                                                 
  cmp   r1, r0                                                                 
  bne   .LBB0_2                                                                
@ BB#1:                                                                                
  movs  r3, #1                                                                 
  mov   r0, r3
  b     .LBB0_3                                                                
.LBB0_2:
  mov   r3, r2                                                                 
  mov   r0, r3                                                                 
  movs  r3, #1                                                                 
.LBB0_3:
  bx    lr
\end{lstlisting}
    \end{minipage}
    }
  \end{tcolorbox}
  \vspace{10pt} 
\caption{\label{lst:leak_cr} Two secure program variants of
  Figure~\ref{lst:cr} for ARM Cortex M0}
\end{figure}

\paragraph{Combined Mitigation:}
Embedded devices that handle sensitive data are vulnerable to both
\acp{SCA} and \acp{CRA}.
Mitigating \acp{SCA} and \acp{CRA} in these devices is essential for
protecting sensitive data and the system.
%
%
A low-overhead approach against \ac{CRA} is fine-grained code
diversification, while software mitigations hinder \acp{SCA} in
cryptographic software.
Avoiding diversifying cryptographic libraries may lead to \acp{CRA},
as shown in recent work by \citet{ahmed_methodologies_2020}, where a
\acp{CRA} attack may use gadgets from OpenSSL, a cryptographic
library.
Similarly, diversifying cryptographic code may break software
mitigations against \acp{SCA}, as we show in Section~\ref{ssec:rq2}.
The latter demonstrates that fine-grained code diversification against
\acp{CRA} and software mitigations against \acp{SCA} constitute conflicting
mitigations. Therefore, there is a need for combined approaches that
protect against the combination of these attacks.

Figures.~\ref{lst:leak} and \ref{lst:leak_cr} show two different
machine-code implementations of programs in
Figures~\ref{lst:masked_all} and \ref{lst:cr}, respectively.
Each of these functions includes code-reuse gadgets that end with
instruction
\lstinline[style=armstyle,firstline=27,lastline=38]{bx lr}.
An attacker may select these gadgets to perform a \ac{CRA}.
Generating multiple versions of each program is a form of
diversification that hinders attacks by altering the attacker's
building blocks.
At the same time, these variants should preserve \ac{SCA} mitigations.
For example, the variant in Figure~\ref{fig:xor-seccon-1} is not secure
against \ac{PSC} attacks.
To tackle this problem, we propose \ac{\toolname}, which generates
diverse variants protected against \acp{CRA} that are also secure against
\acp{SCA}.

\subsection{Threat Model}
\label{ssec:threatmodel}

We assume that the code implementation contains a memory vulnerability
that allows the attacker to perform a \ac{CRA}, in particular a static
\ac{ROP} or \ac{JOP} attack.
We further assume that the attacker does not have direct access to the
memory of the device.
We consider two types of attacker models for \acp{SCA},
\timingattacker that measures the execution time of the
  program and \powerattacker that records the power consumption of the
  program:

\begin{description}
\item[\timingattacker:] The attacker has access to the software
  implementation and the \textit{public} data but not the
  \textit{secret} data.
  The attacker extracts information about the secret data
  by measuring the execution time of the code on the target device.
  The measurements are done remotely.
  
\item[\powerattacker:] The attacker has access to the software
  implementation and the \textit{public} data but not the
  \textit{secret} data.
  At every execution, the program under execution generates new
  \textit{random} values and the attacker has no knowledge of these
  values.
  The attacker extracts information about the secret data by measuring
  the power consumption of the device that the code runs on.
  The attacker may accumulate a number of power traces from multiple
  runs of the program and perform statistical analysis, such as
  \ac{DPA}~\cite{kocher_differential_1999} or
  \ac{CPA}~\cite{brier_correlation_2004-1,papagiannopoulos_mind_2017}.
\end{description}
%
We adopt the leakage model for \acp{PSC} from
\citet{tsoupidi2021seccon} and the leakage model for the
\ac{CR}-policy from \citet{barthe_secure_2021}.

\subsection{Problem Statement}
\label{ssec:pstatement}

Our goal is to generate code secure against the attacker models
\timingattacker and \powerattacker.
First, we define formally  code diversification.
We consider a program $p$ and a set, $S$, of program implementations,
$p_i \in S$, that are functionally equivalent ($\sim$) with the
original program, i.e.\ $\forall p_i \in S. p \sim p_i$ and each other,
$\forall p_i, p_j \in S. p_i \sim p_j$.
To protect against \acp{SCA}, we define a set of constraints $C_{sec}$.
A program implementation $p_i$ is secure against \acp{SCA} (\ac{PSC}
or \ac{TSC} attacks) if $p_i \in sol(C_{sec})$.

To protect small embedded devices that are vulnerable to \acp{CRA} and
\acp{SCA}, \ac{\toolname} generates a pool of diverse solutions,
$S_{sec}$, that is a subset of $S$, and all solutions are secure
against \acp{SCA}, namely they satisfy $C_{sec}$, or $p_i, p_j \in
S_{sec} \subseteq S \implies p_i,p_j \in sol(C_{sec}) \land p_i \sim p_j$,
%
%
The goal of \ac{\toolname} is to generate set $S_{sec}$.

\section{\ac{\toolname}}
\label{sec:approach}
\begin{figure}
  \centering  
  \input{figs/divcon3}
  \caption{\label{fig:secdivcon} High-level view of \ac{\toolname}}
\end{figure}

\ac{\toolname} uses a combinatorial compiler backend to combine
\ac{SCA} mitigations with code diversification against \acp{CRA}.
Figure~\ref{fig:secdivcon} shows a high-level view of \ac{\toolname}.
The input to \ac{\toolname} is 1) the security policy, namely which
input values are \texttt{secret}, \texttt{public}, or \texttt{random},
and 2) the input function in a low-level intermediate representation
generated by a general purpose compiler frontend (CF).
%
%

The first stage of \ac{\toolname} is the \ac{SA} module
(Section~\ref{sec:cr}), which performs code analysis and generates the
input data for the second stage, SecSolver (Section~\ref{ssec:sm}).
SecSolver solves the \ac{SCA}-aware constraint-based backend model
and generates the best-found solution according to the cost function.
Then, SecSolver passes this solution together with the constraint
model to the third stage, SecDiv (Section~\ref{ssec:secdiv}), a
constraint-based diversification method that generates multiple
\ac{SCA}-aware solutions.
%
%
The following sections describe each of the stages of \ac{\toolname}.

\subsection{\acl{SA} Module}
\label{sec:cr}

The \ac{SA} module takes as input the security policy and the input
function (Section~\ref{ssec:sp}).
Subsequently, \ac{\toolname} propagates the security policy to each
program term using type inference (Section~\ref{ssec:ti}).
In cases when the input program is not secure against \acp{SCA},
\ac{\toolname} performs transformations (Section~\ref{ssec:ct}) that
enable the generation of secure code.
The output of the analysis is the extended constraint model of the
input program, which includes data that is necessary for the security
constraints (Section~\ref{ssec:sm}).

%


\subsubsection{Security Policy}
\label{ssec:sp}

\ac{\toolname} takes as input a filename that defines the
  security policy of the function under analysis.
  In particular, the security policy defines the security type of
  each function argument.
  For example, in Figure~\ref{lst:masked_all}, we want to define
  that 1) the data of the first argument is known to the attacker
  (\lstinline[style=spstyle, numbers=none]{Public}), 2) the
  second argument contains secret information
  (\lstinline[style=spstyle, numbers=none]{Secret}), and 3) the third
  argument is a randomly generated value (\lstinline[style=spstyle,
    numbers=none]{Random}).
  \ac{\toolname} takes as input this information in a file as
  follows:

  \begin{tcolorbox}[colback=\codebackcolor, colframe=\codebackcolor,
    top=-8pt, bottom = -8pt, right = 0pt, left = 0pt]
\begin{lstlisting}[style=spstyle, numbers=none]
Public r0;
Secret r1;
Random r2

\end{lstlisting}
  \end{tcolorbox}

  Given the definition of the types for the input arguments,
  \ac{\toolname} implements a type-inference algorithm at a low-level
  intermediate representation format to extract explicit timing leaks
  and transitional power leaks. 

\subsubsection{Type Inference}
\label{ssec:ti}
For both attacker models, \timingattacker and \powerattacker,
\ac{\toolname} uses type inference to propagate a type,
i.e.\ \texttt{secret}, \texttt{public}, or \texttt{random}, to each
program variable.
For example, in Figure~\ref{lst:cr}, intermediate variable \texttt{t}
takes the type \texttt{public}.
Similarly, in Figure~\ref{lst:masked}, intermediate variables
\texttt{mk} and \texttt{t} are assigned type \texttt{random} (because
\texttt{mask} randomizes the value of \texttt{key}).
Software masking introduces additional challenges to the type
inference algorithm, which has to capture properties such as
$(\texttt{sec} \oplus \texttt{mask}) \oplus \texttt{mask} =
\texttt{sec}$.
To achieve this, the inference algorithm uses additional environment
structures that keep track of the random and secret values that an
intermediate variable may contain.
The type-inference algorithm that considers random values is based on
previous work~\cite{gao_verifying_2019-1,wang_mitigating_2019-1}.

\subsubsection{Code Transformations}
\label{ssec:ct}
The implementations of C or C++ programs that are given as input to
\ac{\toolname} may not be secure.
Furthermore, the general-purpose middle-end compiler transformations that
\ac{\toolname} uses may break some of the high-level mitigations.
%
In particular, \ac{\toolname} needs to preserve the \ac{CR} property
against \timingattacker.
However, some of the secret-dependent branches may not be balanced
in the source code, or the high-level compiler optimizations may
remove dead basic blocks~\cite{dsilva_correctness-security_2015-1}.
Similarly, \ac{\toolname} needs to generate secure masked code against
\powerattacker.
The input code is masked, however, high-level optimizations are known
to invalidate some masking
countermeasures~\cite{athanasiou_automatic_2020-1,tsoupidi2021seccon}.
In the following paragraphs, we discuss the program transformations
that \ac{\toolname} implements before the solving stage protecting against
\timingattacker or \powerattacker.

\paragraph{Timing Attacker (\timingattacker):}

\ac{CR} programs may contain secret-dependent branches. However, these
branches should not result in any execution-time differences.
Yet, sometimes, the source code of the input program contains
unbalanced secret-dependent branches.
Figure~\ref{lst:cr} shows a program that branches on the secret value
(line 3).
If the condition is \texttt{true}, the execution takes at least one
cycle (line 4), whereas if the condition is \texttt{false}, it takes
zero cycles.
To deal with these programs, we introduce a transformation that
balances secret-dependent blocks, using two methods: 1) \ac{EBB},
which inserts an empty block, and 2) \ac{CBB}, which copies the
present secret-dependent block in an \texttt{else} statement.

\begin{description}
\item[\acs{EBB}:]
  To balance an unbalanced secret-dependent block, EBB adds an empty
  block that contains \ac{NOP} operations.
  Figure~\ref{lst:empty} shows a
  secret-dependent branch that is balanced using an empty basic
  block (lines 5-7).
  At a later stage, the constraint solver fills this basic block with
  an appropriate number of \ac{NOP} instructions to balance the
  secret-dependent branch.
  In contrast to \ac{CBB}, this transformation works also when we want to
  balance an unbalanced path with more than one basic blocks.
\item[\acs{CBB}:]
  Another way to balance a secret-dependent block that consists of one
  basic block is by copying the unbalanced block instructions.
  Figure~\ref{lst:copy} shows a secret-dependent branch, where the
  else branch is a copy of the if-branch body with inactive
  instructions.
  Here, \ac{\toolname} copies the body of the secret-dependent
  branch to a new \texttt{else} body, which contains all the
  operations of the original block but assigned to unused variables
  (lines 5-7).
\end{description}

\begin{figure}
\centering
    \begin{tcolorbox}[colback=\codebackcolor, colframe=\codebackcolor,
      top=-8pt, bottom = -18pt,
      right = 0pt, left = 0pt]
\subfloat[Add Empty Block\label{lst:empty}]{
    \begin{minipage}[b]{0.48\textwidth}
\begin{lstlisting}[style=cstyle]
u32 check_bit(u32 pub, u32 key) {
  u32 t = 0;
  if (pub == key)
    t = 1;
  else
    // nop;
  return t;
}
\end{lstlisting}
    \vspace{-5pt}
    \end{minipage}
}
    \subfloat[Copy Unbalanced Block\label{lst:copy}]{
      \begin{minipage}[b]{0.48\textwidth}
\begin{lstlisting}[style=cstyle, firstnumber=2, firstline=2]
  u32 _t, t = 0;
  if (pub == key)
    t = 1;
  else
    _t = 1;
  return t;
}
\end{lstlisting}
      \vspace{-5pt}
      \end{minipage}
    }
  \end{tcolorbox}
  \vspace{10pt}
\caption{\label{lst:mitigations} Balancing transformations for
  Figure~\ref{lst:cr}}
\end{figure}

\paragraph{Power Attacker (\powerattacker):}

%
Previous work has shown that high-level compiler optimizations may
break software masking against \ac{PSC}
attacks~\cite{athanasiou_automatic_2020-1,tsoupidi2021seccon}.
For example, Figure~\ref{lst:unxor} shows the result of high-level
compiler optimizations (-O1 to -O3) on masked code.
The code performs first the xor operation between the
public value \texttt{pub} and the secret value \texttt{key} (line 2)
and then, performs the second xor operation with the result of
the first and the \texttt{mask}.
Performing the operations in this order fails to randomize the secret
value and leads to a \ac{PSC} leak at line 2.
To mitigate this type of transformation, \ac{\toolname} transforms the
code to the original operand order (see Figure~\ref{lst:masked}).


\subsection{Security Constraint Model}
\label{ssec:sm}

%
SecSolv (see Figure~\ref{fig:secdivcon}) takes as input the data from
the \ac{SA} module and applies constraints in order to generate
\ac{SCA}-aware code.
First, we will give an overview of a combinatorial compiler backend
(Section~\ref{ssec:cpmodel}) and then proceed with the \ac{SCA}-aware
model (Section~\ref{sec:scaconstr}).

\subsubsection{Constraint-based Compiler Backend}
\label{ssec:cpmodel}
%
We consider a constraint-based compiler backend that implements two
low-level optimizations, instruction scheduling and register
allocation~\cite{lozano_survey_2019}.
A constraint model defines all legal instruction orders and register
assignments~\cite{lozano_combinatorial_2019}.
More formally, a constraint-based compiler backend may be modeled as a
\ac{COP}, $P =\langle V, U, C, O\rangle$, where $V$ is the set of
decision variables of the problem, $U$ is the domain of these
variables, $C$ is the set of constraints among the variables, and $O$
is the objective function.
A constraint-based compiler backend aims at minimizing $O$, which
typically models the execution time or size of the code.

A program is modeled as a set of basic blocks $B$.
Each basic block contains a number of optional operations that may be
\textit{active} or not.
An active operation appears in the final generated binary
  code, whereas inactive operations are not present in the final
  code.
A set of hardware instructions may implement each operation that
consists of a number of operands.
Each operand may be implemented by different, equally-valued virtual
registers, which are the result of copying the content of a
  register to another register or memory (copies).
The model maps each virtual register to a set of hardware registers
and memory locations.
The solver assigns each virtual register with one hardware register
or memory location.
Every assignment $p$ of the problemc variables that satisfies the
constraints, $C$, is a solution to $P$, $p \in sol(P)$ and represents
a compiled program.


A typical objective function of a constraint-based backend minimizes
different metrics such as \textit{code size} and \textit{execution time}.
These can be captured in a generic objective function that sums up
the weighted cost of each basic block:
\[
\sum_{b\in B} \mathit{weight}(b) \cdot \textbf{cost}(b).
\]
The \textbf{cost} of each basic block is a variable that differs among
solutions, whereas \textit{weight} is a constant value that represents
the contribution of the specific basic block to the total cost.
This cost model is accurate for predictable hardware
architectures, such as microcontrollers.
These architectures do not include cache hierarchy, dynamic
  branch prediction, and/or out-of-order execution, which reduce
  predictability.

\subsubsection{Side-Channel Mitigation Constraints}
\label{sec:scaconstr}

The constraint-based solver aims at optimizing code given an accurate
cost model for predictable microcontrollers.
However, \ac{\toolname} aims at generating \ac{SCA}-secure code.
Given the constraint problem $P =\langle V, U, C, O\rangle$ that
describes the combinatorial compiler backend, we extend the
constraints $C$, with a set of constraints $C_{sec}$ that capture the
properties of the \ac{SCA} mitigations.
Then, the problem becomes $P_{sec} =\langle V, U, C \cup C_{sec},
O\rangle$ and the goal is to find the solution that optimizes the cost
function, $O$, while satisfying all constraints.
The following paragraphs describe briefly the constraints for the two
attacker models.

\paragraph{Timing Attacker (\timingattacker):}
For \timingattacker, the \ac{SA} module generates a list of sets of
paths, $paths_{sec}$.
Each element in the list contains the set of possible paths starting
from one secret-dependent branch.
To generate the set of paths \ac{SA} applies a path-finding algorithm
(see Section~\ref{app:algorithm}). 
%

The constraints that guarantee the preservation of the \ac{CR} policy
are based on the paths ($paths_{sec}$) that \ac{SA} provides to the
solver.
For each set of paths that depends on a secret value, we define a
constraint \texttt{balance\_blocks}, which guaranties that all paths
in the set have the same execution time.

\begin{tcolorbox}[colback=\colbackcolor,colframe=\colframecolor, top=-8pt, bottom = -8pt,
  right = 0pt, left = 0pt]
  \begin{lstlisting}[style = modelingstyle]
balance_blocks($paths_{sec}$):
   $\forall p_1, p_2 \in paths_{sec} . \sum_{b\in p_1} \texttt{cost}(b) =
                                     \sum_{b\in p_2} \texttt{cost}(b)$
\end{lstlisting}
\end{tcolorbox}

In particular, for each set of paths that depend on a \texttt{secret}
value ($sec_i$), we apply the \texttt{balance\_blocks} constraint,
i.e.\ $\forall paths_{sec_{i}} \in \allowbreak
paths_{sec}~.\allowbreak~\texttt{balance\_blocks}(paths_{sec_{i}})$.
In this case, we have one security constraint, i.e.\ $C_{sec} =
\{\texttt{balance\_blocks}\}$.


\paragraph{Power Attacker (\powerattacker):}

The model against \acp{PSC} depends on the constraint model in
previous work~\cite{tsoupidi2021seccon}.
This model focuses on two leakage sources, namely, \ac{ROT} and
\ac{MRE}.

\ac{ROT} leakages occur when there is a value transition in a
hardware register, namely when a new value replaces the previous value
of a register.
When this transition depends on a secret value, we have a secret leak.
The constraint model enforces the absence of these leaks in the
generated code by constraining register allocation.
More specifically, for \ac{ROT} leaks, \ac{SA} generates all pairs of
intermediate variables, $(t_1,t_2) \in RPairs$, that should not be
assigned to the same register ($r(t)$) subsequently (\texttt{subseq}).

\begin{tcolorbox}[colback=\colbackcolor,colframe=\colframecolor, top=-8pt, bottom = -8pt,
  right = 0pt, left = 0pt]
  \begin{lstlisting}[style = modelingstyle]
conflict_rassign($RPairs$):
  $\forall \texttt{t}_1, \texttt{t}_2 \in
RPairs .~ r(\texttt{t}_1) = r(\texttt{t}_2) \implies \neg \texttt{subseq}(\texttt{t}_1,\texttt{t}_2) $
\end{lstlisting}
\end{tcolorbox}

Similarly, \ac{MRE} corresponds to a leak when there is a
secret-dependent transition at the memory bus, i.e.\ when a load or
store operation overwrites the previous value in the bus.
The constraints that ensure secure code generation are similar with
those against \ac{ROT} and enforce the instruction order of memory
operations.
In particular, for \ac{MRE} leaks, \ac{SA} generates ~~ all ~~~~pairs ~~of
~~~~memory ~~operations, ~~~~ $(o_1,o_2) \in MPairs$, that should not be
scheduled one after the other (\texttt{msubseq}).
\begin{tcolorbox}[colback=\colbackcolor,colframe=\colframecolor, top=-8pt, bottom = -8pt,
  right = 0pt, left = 0pt]
  \begin{lstlisting}[style = modelingstyle]
conflict_order($MPairs$):
        $\forall \texttt{o}_1, \texttt{o}_2 \in
          MPairs .~ \neg \texttt{msubseq}(\texttt{o}_1,\texttt{o}_2)$
\end{lstlisting}
\end{tcolorbox}

In this case, we have two security constraints that protect against
\ac{ROT} and \ac{MRE} leaks, i.e.\ $C_{sec} =\allowbreak
\{\texttt{conflict\_rassign},\allowbreak \texttt{conflict\_order}\}$.



\subsection{Secure Code Diversification}
\label{ssec:secdiv}
Constraint-based
diversification~\cite{hebrard_distance_2007,ingmar_modelling_2020}
aims at generating different solutions for a given problem rather than
one solution.
For optimization problems, there is often the requirement to generate
good or optimal solutions with regard to the optimality function,
$O$.
Constraint-based diversification defines the notion of
\textit{distance measure}, $\delta$, which is a constraint between
problem solutions and measures how \textit{different} two solutions of
the problem are.

\subsubsection{Diversification Problem}
Given our \ac{SCA}-aware optimization problem $P_{sec} =\langle V, U,
C \cup C_{sec}, O\rangle$, the diversification problem attempts to
find the set of distinct solutions $S$ that are solutions of $P'_{sec}
=\langle V, U, C\cup C_{sec}\rangle$ and the distance between the
solutions satisfies $\delta$, i.e.\ $S = \{p \,|\, p \in sol(P'_{sec})
\land \forall p' \in S\,.\, p'\neq p \implies \delta(p,p')\}$.
%


To generate the set of diverse programs $S$, SecDiv (see
Figure~\ref{fig:secdivcon}) takes the best solution from the code
generation part and generates multiple solutions using the
security-aware constraint model (SecSolver in
Figure~\ref{fig:secdivcon}), similar to previous
work~\cite{tsoupidi2021constraint}.
In particular, SecDiv generates solutions in the ~~~neighbourhood ~~of ~~this
~~solution ~~~~that ~~satisfy $C_{opt} = O < g \cdot o$, where $g$ is the
maximum allowed optimality gap and $o$ the cost of the best-found
solution.

\subsubsection{Diversifying Transformations}
  The diversifying transformations that SecDiv supports are 1)
  hardware register assignment, 2) register copying, 3) memory
  spilling, 4) constant rematerialization, 5) instruction order, 6) \ac{NOP}
  insertion, and 7) operand order in two-address instructions.
  Hardware register assignment permits changing the register assignment
  for instruction operands.
  Register copying enables copying the content of a register to another
  register for future uses of the register value.
  Memory spilling allows copying values from a register to the stack and
  from the stack to a register.
  Spilling affects the size of the stack and thus leads to stack size
  diversification, however, spilling increases execution-time overhead.
  Rematerialization allows re-executing an instruction instead of
  copying its result.
  SecDiv may also alter the instruction order as long as there are not
  data dependencies and insert \ac{NOP} instructions by delaying the
  issue cycle of an instruction.
  Finally, SecDiv may alter the operand order in two-address
  instructions.

  \section{Evaluation}
\label{sec:eval}
The evaluation consists of two parts: 1) the identification
  of side-channel vulnerabilities in a diversification tool and 2) the
  evaluation of \ac{\toolname}.
The evaluation of \ac{\toolname} focuses on two aspects: a) the
effectiveness of the \ac{CR}-preserving code generation and b) the
effectiveness of \ac{SCA}-aware code diversification.
Section~\ref{sec:preliminaries} describes the implementation of
\ac{\toolname}, experimental setup, and benchmarks, while
Section~\ref{ssec:rq2} presents our findings with regard to \ac{SCA}
vulnerabilities in diversified code.
Sections~\ref{ssec:rq1}-\ref{ssec:rq4} present the evaluation of
\ac{\toolname}.

\subsection{Evaluation Setup}
\label{sec:preliminaries}
In the following parts, we present the implementation, experimental
setup, and benchmarks we use to evaluate \ac{\toolname}.

\subsubsection{Implementation}

We implement \ac{\toolname} as an extension of
Unison~\cite{lozano_combinatorial_2019}, a combinatorial compiler
backend that uses \ac{CP}~\cite{rossi2006handbook} to optimize
software functions.
To do this, Unison combines two low-level optimizations, instruction
scheduling and register allocation, and achieves optimizing
medium-size functions with improvement over
LLVM~\cite{lozano_combinatorial_2019}.
\ac{\toolname} takes as input the function in LLVM's \ac{MIR} and
outputs multiple versions of the function that satisfy the compiler
and security constraints.
For generating \ac{PSC}-free code, we adapt the model of
SecCG~\cite{tsoupidi2021seccon}, which generates optimal code that is
secure against \ac{ROT} and \ac{MRE} leakages.
For generating \ac{CR} code, we implement the path-extraction
algorithm (see Section~\ref{app:algorithm}) in Haskell as
part of Unison's presolving process.
We implement the path-balancing constraints (see
Section~\ref{sec:scaconstr}) as part of the constraint model, which is
written using the Gecode C++ library~\cite{Gecode2020}.
\ac{\toolname} combines the \ac{SCA}-aware mitigations with a
diversification scheme~\cite{tsoupidi2021constraint} to generate
multiple function variants.
We target two architectures, 1) a generic MIPS32 processor and 2) the
ARM Cortex M0 processor~\cite{armcortex}.

\subsubsection{Experimental Setup}
\label{ssec:expsetup}
All experiments run on an Intel%
\textsuperscript{\textregistered}%
Core\texttrademark i9-9920X processor with maximum frequency 3.50GHz
per core and 64 GB of RAM running Debian GNU/Linux 10 (buster).
We use LLVM-3.8 as the front-end and middle-end compiler for these
experiments.
We repeat all experiments five times, with different random seeds
(where applicable) and report the mean value for each metric in the
results.

\subsubsection{Benchmarks}
\label{ssec:benchs}

\begin{table}[h]
  \centering
  \caption{\label{tab:bench_description} Benchmark description; $N_i$
    is the number of machine instructions that are input to the
    compiler backend; $N_b$ is the number of basic blocks; A stands
    for ARM Cortex M0; and M stands for MIPS32; $I_p$, $I_s$, and $I_r$ is the
      number of public, secret, and random input
      arguments, respectively}
  \setlength{\tabcolsep}{3.2pt}
  \begin{tabular}{|l|l|r|r|r|r|r|r|r|}
    \hline
    \multirow{2}{*}{Prg} & \multirow{2}{*}{Description} & \multicolumn{2}{c|}{$N_i$} &\multicolumn{2}{c|}{$N_b$} & \multicolumn{3}{c|}{Input Vars} \\\cline{3-9}
    &   & A & M & A & M & I$_{p}$ & I$_{s}$ & I$_{r}$\\\hline
    P0  & SecXor                        & 7   & 7  &1 &1 & 1  & 1 &  1  \\
    P1  & AES Shift Rows               & 8   & 8  &1 &1 & 0  & 2 &  2  \\
    P2  & Messerges Boolean            & 11  & 11 &1 &1 & 0  & 1 &  2  \\
    P3  & Goubin Boolean               & 13  & 13 &1 &1 & 0  & 1 &  2  \\
    P4  & SecMultOpt\_wires          & 18  & 18 &1 &1 & 1  & 1 &  3  \\
    P5  & SecMult\_wires             & 18  & 18 &1 &1 & 1  & 1 &  3  \\
    P6  & SecMultLinear\_wires       & 19  & 19 &1 &1 & 1  & 1 &  3  \\
    P7  & CPRR13-lut\_wires       & 48  & 48 &1 &1 & 1  & 1 &  7  \\
    P8  & CPRR13-OptLUT\_wires    & 48  & 48 &1 &1 & 1  & 1 &  7  \\
    P9  & CPRR13-1\_wires         & 52  & 52 &1 &1 & 1  & 1 &  7  \\
    P10  & Whitening                   & 113 & 88 &1 &1 & 16 &16 & 16  \\
    \hline                                                                         
    C0  & If check (Figure~\ref{lst:cr})  & 10  & 9  &3 &3 & 1  & 1 &  -  \\
    C1  & Share's Value          & 23  & 26 &6 &5 & 1$^{\mathrm{a}}$ & 2$^{\mathrm{a}}$ &  -\\
    C2  & Mult. Modulo 8         & 28  & 24 &8 &6 & 1  & 1  &  - \\
    C3  & Modulo Exponentiation  & 51  & 36 &7 &7 & 1  & 2  &  - \\
    C4  & Kruskal                & 51  & 55 &9 &9 & 1$^{\mathrm{a}}$ & 3$^{\mathrm{a}}$ &  -\\
    \hline
    \multicolumn{9}{l}{$^{\mathrm{a}}$The input is an address to an array of secret values}
  \end{tabular}
\end{table}


Our approach concerns programs that handle secret information and are,
thus, vulnerable to \acp{SCA}.
Therefore, we have selected eleven masked cryptographic core functions
that may be vulnerable to
\acp{PSC}~\cite{tsoupidi2021seccon,wang_mitigating_2019-1} and five
functions that exhibit secret-dependent timing variations and are used
in cryptographic context~\cite{mantel_transforming_2015}.
Table~\ref{tab:bench_description} shows the benchmarks with
information about the origin of the function, the function size in
number of instructions ($N_i$) and the number of basic blocks ($N_b$)
for ARM Thumb (A) and MIPS32 (M), and finally, the input variables,
($I_p$ public, $I_s$ secret, and $I_r$ random input variables).

\paragraph{Masked Programs:}
\label{ssec:maskedprogs}

The masked programs that we use in this evaluation consist of eleven
programs, P0 to P10, most of which originate from the work by
\citet{wang_mitigating_2019-1}.
These benchmark programs consist of masked cryptographic core
functions that may be vulnerable to \ac{PSC} attacks.
More specifically, P0 is a masked exclusive-OR
  implementation, P1-P3 are protected by Boolean
  masking~\cite{bayrak_sleuth_2013}, P4-P6 correspond to masked
  multiplication~\cite{rivain_provably_2010}, P7-P9 correspond to
  masked S-box for AES~\cite{coron_higher-order_2014}, and P10
  implements key whitening ~\cite{bayrak_sleuth_2013}.

\paragraph{\ac{CR} Programs:}
For evaluating the \ac{CR} property we use Listing~\ref{lst:cr} and
four benchmark programs used by \citet{mantel_transforming_2015}.
The code for these benchmarks in C including security-policy
annotations is available by \citet{winderix_compiler-assisted_2021}.
  The set of benchmark programs includes 1) C0 is simple unbalanced
  \texttt{if} statement (see Figure~\ref{lst:cr}), 2) C1 computes the
  total market value of a share from the portfolio of the user, 3) C2
  implements multiplication modulo 8, 4) C3 implements the
  square-and-multiply modular exponentiation, and 5) C4 implements
  Kruskal's algorithm~\cite{kruskal_shortest_1956} to compute the
  minimum spanning tree of a graph.
These implementations are vulnerable to timing
attacks~\cite{mantel_transforming_2015}.

  \subsection{Side-Channel Vulnerabilities in Code Diversification}
\label{ssec:rq2}

Our approach is based on the intuition that automatic
  software diversification transformations may break side-channel
  mitigations.
  To confirm this intuition, we investigate the effect of code
  diversification using a freely-available\footnote{MCR:
    \url{https://github.com/securesystemslab/multicompiler.git}} code
  diversification tool, \ac{MCR}~\cite{homescu_profile-guided_2013}.
  In particular, we analyze code variants generated by \ac{MCR} to
  check if the code violates software mitigations against \acp{SCA}.
  To do that, we use \ac{MCR} to diversify benchmark programs that
implement security mitigations against \acp{SCA} at source-code level.
Then we verify whether the generated program variants (for the
respective benchmarks) satisfy the software mitigations against
\ac{PSC} or \ac{TSC} attacks.
For \ac{PSC}, we use a tool\footnote{FSE19 tool:
  \url{https://github.com/bobowang2333/FSE19}} by
\citet{wang_mitigating_2019-1}, whereas for \ac{TSC}, we measure the
execution time manually.
Both tools focus on the x86 architecture.
%
For these experiments, we generate 50 random variants by providing 50
different random seeds to \ac{MCR}.
\ac{MCR} supports randomization at multiple layers of the compilation
process, including hardware-register randomization and
\ac{NOP}-insertion.
These randomizing transformations may affect \ac{PSC} and \ac{TSC}
mitigations, respectively.
In the following paragraphs, we investigate how hardware-register
randomization affects \ac{ROT} leakages and how \ac{NOP}-insertion
affects the \ac{CR} property.

\paragraph{Hardware-Register Randomization:}
Hardware-register randomization~\cite{crane_readactor_2015} is a form
of fine-grained software diversification that generates program
variants that differ with regard to the register assignment at the
register-allocation stage of the compilation process.
Among other transformations, \ac{MCR} implements hardware-register
randomization.
To identify the number of \ac{ROT} leaks of each of the variants, we
implement parts of the tool by \citet{wang_mitigating_2019-1} to
extract information from the register allocation step in \ac{MCR}.
Subsequently, we use the tool by \citet{wang_mitigating_2019-1}
to identify the leaks in the variants.

For each of the masked benchmarks, Table~\ref{tab:multi} shows the
number of leaks that appear in the baseline, which uses the LLVM
compiler~\cite{wang_mitigating_2019-1} and the rate of variants that
contain different numbers of leaks after diversification with
\ac{MCR}.
The last column shows the rate of variants that have at least one
leak.
Overall, there are leaking variants in all programs, ranging from 62\%
for P0 and 100\% for P2, P4, P6-P10.
For programs P0 to P6, the number of leaks differs for the generated
variant population.
In particular, \ac{MCR} may introduce leaks in P1, P2, and P6 that the
baseline does not generate.
Inversely, \ac{MCR} may generate variants that are leak-free for P0,
P1, P3, and P5.
This means that the hardware-register randomization transformation
allows the generation of leak-free variants.

To summarize, we observe that randomization may break masking
mitigations, whereas, in many cases, there is a space for generating
leak-free variants.


\begin{table}[h]
  \centering
  \caption{\label{tab:multi} \ac{ROT} vulnerabilities in diversified code by \acf{MCR}}
  \begin{tabular}{|l|r|r|r|}
    \hline
    \multirow{2}{*}{Prg} & \cite{wang_mitigating_2019-1} & \multicolumn{2}{c|}{MCR} \\\cline{2-4}
    & \#leaks     & \#leaks (\% of variants)   &  $\ge$ one leak\\\hline
    P0     & 1    & 1 (62\%) 0 (38\%)          &  62\% \\
    P1     & 0    & 2 (92\%) 1 (2\%) 0 (6\%)   &  94\%  \\
    P2     & 1    & 2 (100\%)                  &  100\% \\
    P3     & 1    & 1 (70\%) 0 (30\%)          &  70\%  \\
    P4     & 1    & 1 (100\%)                  &  100\% \\
    P5     & 1    & 1 (96\%) 0 (4\%)           &  96\%  \\
    P6     & 3    & 4 (52\%) 5 (48\%)          &  100\% \\
    P7     & 14   & 14 (100\%)                 &  100\% \\
    P8     & 16   & 16 (100\%)                 &  100\% \\
    P9     & 12   & 12 (100\%)                 &  100\% \\
    P10    & 5    & 5 (100\%)                  &  100\% \\
    \hline
  \end{tabular}
\end{table}

\paragraph{\ac{NOP} Insertion:}

\ac{NOP} insertion is a form of fine-grained software diversification
that generates program variants that contain randomly inserted
\ac{NOP} operations.
\ac{MCR} implements \ac{NOP}-insertion
randomization~\cite{homescu_profile-guided_2013}.
The source code of programs C0 to C3 does not comply with the \ac{CR}
policy.
To identify \ac{CR} violations, we consider C0, C1 and C3 because they
are simple to verify manually.
We modify the C implementations of C0, C1 and C3 to balance the
secret-dependent branches and consider a simple timing model for the
processor that considers one cycle per instruction.
The results are that 88\% of C0, 74\% of C1, and 72\% of C3 are
unbalanced.
\ac{MCR} inserts \ac{NOP} operations randomly without information
about secret balancing and, thus, generates non-\ac{CR}-preserving
code\footnote{Here, we do not investigate multi-variant execution,
  where different variants are loaded dynamically, which may hinder
  timing attacks by randomizing the execution time.}.
To summarize, \ac{NOP} insertion may break branch balancing for \ac{CR}
programs.

\subsection{Effectiveness and Efficiency of \acs{TSC}-Aware Code Generation}
\label{ssec:rq1}

\begin{table}[h]
  \centering
  \caption{\label{tab:cr} Optimality overhead in cycles for
    \ac{CR}-preserving code-generation; \secured denotes secure
    variants and \insecured non-secure variants; OH stands for
    Overhead}
      \setlength{\tabcolsep}{5pt}
  \begin{tabular}{|l|rr|r|rr|r|}
    \hline
    \multirow{3}{*}{Prg} & \multicolumn{3}{c|}{ARM Cortex M0} & \multicolumn{3}{c|}{Mips32} \\\cline{2-7}
    &\multicolumn{2}{c|}{Cycles} & \multirow{2}{*}{OH (\%)} & \multicolumn{2}{c|}{Cycles}& \multirow{2}{*}{OH (\%)}\\\cline{2-3}\cline{5-6}
    &\secured& \insecured &  & \secured & \insecured& \\\hline 
C0	&	26	&	20	&	30	&	13	&	10	&	30\\
C1	&	1406	&	1220	&	15	&	1105	&	857	&	28\\
C2	&	1039	&	803	&	29	&	975	&	571	&	70\\
C3	&	3012	&	1984	&	51	&	7641	&	5843	&	30\\
C4	&	16130	&	13590	&	18	&	10429	&	8905	&	17\\
\hline
  \end{tabular}
\end{table}

\begin{table}[h]
    \centering
  \caption{\label{tab:cr_stime} Compilation-time overhead in seconds
    for \ac{CR}-preserving code generation; \secured denotes secure
    variants and \insecured non-secure variants; OH stands for
    Overhead}
      \setlength{\tabcolsep}{5pt}
  \begin{tabular}{|l|rr|r|rr|r|}
    \hline
    \multirow{3}{*}{Prg} & \multicolumn{3}{c|}{ARM Cortex M0} & \multicolumn{3}{c|}{Mips32} \\\cline{2-7}
    &\multicolumn{2}{c|}{t (s)} & \multirow{2}{*}{OH (\%)} & \multicolumn{2}{c|}{t(s)}& \multirow{2}{*}{OH (\%)}\\\cline{2-3}\cline{5-6}
    &\secured& \insecured &  & \secured & \insecured& \\\hline 
C0	&	0.32	&	0.19	&	68	&	0.81	&	0.55	&	47\\
C1	&	1.68	&	0.91	&	84	&	4.42	&	2.56	&	72\\
C2	&	8.48	&	0.81	&	946	&	2.65	&	1.33	&	99\\
C3	&	57.26	&	23.46	&	144	&	8.02	&	4.97	&	61\\
C4	&	150.80	&	92.72	&	62	&	27.52	&	7.93	&	247\\
\hline
  \end{tabular}
\end{table}

\begin{table}[h]
\centering
  \caption{\label{tab:sec} Security Evaluation using a WCET tool to
    compare the execution time: $\top$ denotes a symbolic value, $v$
    denotes a set of concrete values, and $a_i$ corresponds to the
    $i_{th}$ input argument } \setlength{\tabcolsep}{5pt}
\begin{tabular}{|l|p{2.4cm}|r|p{2.9cm}|r|}
    \hline
    \multirow{2}{*}{Prg} & \multicolumn{2}{c|}{ARM Cortex M0} & \multicolumn{2}{c|}{Mips32} \\\cline{2-5}
    &Input &\secured & Input& \secured\\\hline
C0	&a0,a1 =$\top$&	\cmark	&	a0,a1 = $\top$ &       \cmark\\

C1	&a0,a1,a2,a3 = $\top^{\mathrm{a}}$&	\cmark	&	a0,a1,a3 = $\top$, a2 = $v$     &       \cmark\\
C2	& a0,a1,a2,a3 = $\top$ & \cmark		&	a0,a1,a2,a3 = $\top$            &       \cmark\\
C3	&a0,a1,a2,a3 = $\top^{\mathrm{a}}$  & \cmark		&	a0,a1,a3 = $\top$, a2 = $v$     &       \cmark\\
C4	&a0,a1,a2,a3 = $v^{\mathrm{a,b}}$&	\cmark	&	a0,a1,a2 = $\top$, a3 = $v$     &       \cmark\\
\hline
\multicolumn{5}{p{0.45\textwidth}}{$^{\mathrm{a}}$Verified only the
  secret-dependent branches to improve scalability and accuracy}\\
\multicolumn{5}{p{0.45\textwidth}}{$^{\mathrm{b}}$The concrete values correspond to addresses of the inputs}
  \end{tabular}
\end{table}


%
This section evaluates the \ac{CR}-preserving code generation in three
dimensions, 1) performance overhead, 2) compilation overhead, and 3)
security.

\subsubsection{Performance Overhead}

The \ac{CR}-preserving code generation extends
Unison~\cite{lozano_combinatorial_2019} with constraints that enforce
the \ac{CR} property.
For optimizing code against \acp{TSC}, \ac{\toolname} optimizes the
generated code given the compiler-backend constraints and the newly
introduced security constraints.
Generating \ac{CR}-preserving programs introduces performance overhead
due to the introduction of new basic blocks and/or \ac{NOP} padding
for balancing secret-dependent branches.
To estimate the overhead on the generated code, we utilize the cost
model (see Section~\ref{ssec:cpmodel}) of the constraint-based
compiler backend~\cite{lozano_combinatorial_2019}.
Table~\ref{tab:cr} shows the performance overhead of the
\ac{CR}-preserving code generation backend of \ac{\toolname}
(\secured) compared to Unison that is not security aware (\insecured).
\ac{\toolname} has a maximum overhead of 70\% over Unison for C2.
The introduced overhead is due to the introduction of new basic blocks
and the extension of other basic blocks in order to balance
secret-dependent execution paths.
In contrast to the \ac{CR}-preserving code generation, \acs{PSC}-aware
code generation does not introduce significant execution-time
overhead~\cite{tsoupidi2021seccon}.
%
%
One reason for this is that the \ac{CR} policy affects directly the
execution time of the generated code because it enforces
secret-dependent block balance by increasing the execution time of 
all secret-dependent paths to reach the longest path.

\subsubsection{Compilation Overhead} 
The introdution of new constraints to satisfy the constant-resource
property in the constraint model may lead to increased compilation
time compared to non-secure compilation in Unison.
To evaluate the compilation-time overhead, we compare the compilation
time of \ac{\toolname} with Unison measuring the solving time.
Table~\ref{tab:cr_stime} shows the compilation-time overhead of
\ac{\toolname} (\secured) compared to Unison (non-\ac{CR}-preserving code
optimization)~\cite{lozano_combinatorial_2019} (\insecured).
For ARM Cortex M0, the compilation time is at most ten times
  slower in \ac{\toolname} compared to Unison for C2.
For MIPS, we observe lower slowdown up to 3.5 times for C4.
\ac{PSC}-aware code generation~\cite{tsoupidi2021seccon} demonstrates
a similar difference in the compilation-time slowdown between the two
architectures.
Here, the introduced compilation-time slowdown is mainly due to the
introduced constraints for balancing the cost of different paths,
which introduces inter-block dependencies that delay the solving
process.
At the same time, we notice larger absolute compilation
  times for ARM cortex M0 than for MIPS.
  This is due to the characteristics of the ARM Thumb architecture
  compared to MIPS32, including a smaller number of general-purpose
  hardware registers and two-address instructions.

\subsubsection{Security Evaluation} 
\ac{\toolname} uses a constraint model to generate secure variants.
To verify the effectiveness of \ac{SecDivCon} against timing side
channels, we use two \ac{WCET} tools for the two architectures we are
investigating.
\ac{WCET} is typically a sound overapproximation of the execution time
of the program, whereas \ac{BCET} is a sound underapproximation of the
execution time.
For MIPS, we use KTA~\cite{broman_brief_2017,tsoupidi_two-phase_2017}.
KTA is a tool that extracts the best- and worst-case execution time
for a binary program.
%
%
For evaluating ARM Cortex M0, we use a symbolic-execution-based
\ac{WCET} tool\footnote{CM0 WCET:
  \url{https://github.com/kth-step/HolBA/tree/dev_symbexec_form}} that
generates the \ac{WCET} and \ac{BCET} for a sequence of binary
instructions~\cite{lindner2023proofproducing}.
To verify that \ac{SecDivCon} generates \ac{CR} programs, we test the
generated binaries using a \ac{WCET} tool.
We give as inputs symbolic values that range over all integer values
($\top$) for \texttt{secret} values and \texttt{public} values that do
not affect the control flow, whereas for \texttt{public} values that
affect the control flow (e.g.\ loop bounds), we provide concrete
values.
%
%
%
If the returned \ac{BCET} and \ac{WCET} are equal, then we have
evidence that the program's execution time is secret independent for
the given concrete inputs.
Performing the same experiment using multiple concrete inputs gives an
indication that the program satisfies the \ac{CR} property.
More specifically, we compare
the \ac{WCET} and the \ac{BCET} of the function for different concrete
values of the public inputs.
If $\forall p\in IN_{test}.wcet_p = bcet_p$ for all concrete public
inputs, $IN_{test}$, we say that the program is constant resource
modulo inputs\footnote{Note that possible overapproximations of the \ac{WCET} or
underapproximations of the \ac{BCET} may lead to inequality of
\ac{BCET} and \ac{WCET}, regardless of the program satisfying the
\ac{CR} property.}.
For each of the benchmark programs, Table~\ref{tab:sec} shows the type
of input value we use (Input) and the result of the comparison between
\ac{WCET} and \ac{BCET} (\secured).
For the experiment, we provide different values for the concrete value
$v$.
Symbol \cmark denotes that the experiments for all inputs result in
the same \ac{WCET} and \ac{BCET}.
The result of this experiment indicates that the generated code does
not violate the \ac{CR} property.

\subsection{\ac{SCA}-Mitigation Effect on Code Diversification}
\label{ssec:rq3}

\begin{table*}[h]
  \caption{\label{tab:num_vars} Number of variants (N) and
    diversification time (t) in seconds for \ac{SCA}-aware (\secured)
    and non \ac{SCA}-aware (\insecured) diversification in ARM Cortex
    M0 and Mips32; TO stands for time limit (ten minutes); \ac{\toolname} controls the
    execution-time overhead, here we show the results for a maximum
    execution-time overhead of 0\% and 10\%.}
    \setlength{\tabcolsep}{4.5pt}
  \centering
  \begin{tabular}{|l|a|r|a|r|a|r|a|r|a|r|a|r|a|r|a|r|}
    \hline
    \multirow{4}{*}{Prg} & \multicolumn{8}{c|}{ARM Cortex M0} & \multicolumn{8}{c|}{Mips32} \\\cline{2-17}
        & \multicolumn{4}{c|}{0\%} & \multicolumn{4}{c|}{10\%} & \multicolumn{4}{c|}{0\%} & \multicolumn{4}{c|}{10\%} \\\cline{2-17}
    & \multicolumn{2}{c|}{\secured}& \multicolumn{2}{c|}{\insecured}& \multicolumn{2}{c|}{\secured} & \multicolumn{2}{c|}{\insecured}& \multicolumn{2}{c|}{\secured} & \multicolumn{2}{c|}{\insecured}& \multicolumn{2}{c|}{\secured} & \multicolumn{2}{c|}{\insecured}\\
\hhline{~----------------}
& N & t (s)& N & t (s) & N & t (s) & N & t (s) & N & t (s) & N & t (s) & N & t (s) & N & t (s) \\\hline
P0	&	1	&	-	&	2	&	0.00	&	8	&	8.09	&	18	&	150.88	&	17	&	0.03	&18	&	0.01	&	17	&	0.03	&	18	&	0.01\\
P1	&	5	&	0.17	&	16	&	0.05	&	109	&	194.53	&	200	&	9.47	&	200	&	0.36	&200	&	0.12	&	200	&	1.70	&	200	&	0.26\\
P2	&	2	&	0.00	&	2	&	0.00	&	84	&	397.67	&	65	&	196.98	&	200	&	0.44	&200	&	0.12	&	200	&	3.65	&	200	&	0.41\\
P3	&	39	&	217.16	&	9	&	0.17	&	200	&	73.97	&	200	&	15.90	&	200	&	0.70	&200	&	0.20	&	200	&	4.98	&	200	&	0.51\\
P4	&	200	&	28.83	&	200	&	2.70	&	200	&	27.21	&	200	&	2.09	&	200	&	85.91	&200	&	3.03	&	200	&	74.03	&	200	&	3.80\\
P5	&	200	&	28.76	&	200	&	2.71	&	200	&	27.36	&	200	&	2.10	&	200	&	86.31	&200	&	3.05	&	200	&	73.48	&	200	&	3.78\\
P6	&	200	&	31.18	&	200	&	2.48	&	200	&	29.01	&	200	&	2.25	&	200	&	134.76&200	&	3.75	&	200	&	215.73	&	200	&	3.94\\
P7	&	51	&	TO	&	200	&	18.60	&	51	&	TO	&	200	&	22.69	&	40	&	TO&200	&	20.32	&	32	&	TO	&	200	&	55.32\\
P8	&	69	&	TO	&	200	&	15.47	&	58	&	TO	&	200	&	20.55	&	47	&	TO&200	&	20.40	&	34	&	TO	&	200	&	65.83\\
P9	&	185	&	TO	&	200	&	18.16	&	165	&	TO	&	200	&	23.24	&	6	&	TO&200	&	306.09	&	8	&	TO	&	200	&	171.21\\
P10	&	53	&	TO	&	200	&	23.27	&	20	&	TO	&	200	&	35.28	&	36	&	TO&200	&	16.44	&	15	&	TO	&	200	&	17.34\\
\hline
C0	&	200	&	0.65	&	4	&	41.15	&	200	&	0.38	&	162	&	247.78	&	200	&	0.32	&19	&	0.04	&	200	&	0.46	&	200	&	1.77\\
C1	&	200	&	1.22	&	200	&	1.69	&	200	&	2.96	&	200	&	2.76	&	200	&	0.88	&200	&	0.39	&	200	&	7.66	&	200	&	5.47\\
C2	&	200	&	0.53	&	200	&	0.25	&	200	&	2.51	&	200	&	1.51	&	200	&	0.99	&200	&	0.43	&	200	&	4.43	&	200	&	1.66\\
C3	&	200	&	6.39	&	200	&	5.24	&	200	&	8.55	&	200	&	8.53	&	200	&	4.07	&200	&	2.69	&	200	&	22.09	&	200	&	19.88\\
C4	&	200	&	14.11	&	200	&	10.25	&	200	&	27.53	&	200	&	17.19	&	200	&	10.27	&200	&	7.85	&	200	&	27.87	&	200	&	18.97\\
\hline
  \end{tabular}
\end{table*}

To evaluate the effect of \ac{SCA} mitigations on code
diversification, we compare the effect of \ac{SCA}-aware
diversification with \ac{SCA}-unaware diversification.
We evaluate \ac{\toolname} in two axes, 1) diversity and 2)
diversification scalability.

Table~\ref{tab:num_vars} shows the number of variants (N) and the
diversification time in seconds (t(s)) for each of the
benchmarks and each of the configurations of the diversification
experiments.
The diversification time consists of the time it takes to generate
diverse program variants given an initial optimized solution.
We use a time limit of ten minutes.
In addition, we use upper bound (200) on the number of variants,
because of the increasing complexity of the pairwise gadget-overlap
rate (see Section~\ref{ssec:rq4}) that depends on all pairs of
generated variants.
For each of the two architectures, ARM Cortex M0 and MIPS32, we
perform \ac{SCA}-aware (\secured) diversification and \ac{SCA}-unaware
(\insecured) diversification using 0\% (optimal based on the cost
model) and 10\% optimality gap.
The optimality gap, $p$, depends on the cost model of the
combinatorial compiler backend and the input best-found solution.
The optimality gap results in a constraint that ensures that the cost
of each generated variant is at most $p$\% worse than the best-found
solution.


In the upper part of Table~\ref{tab:num_vars}, we see that for ARM
Thumb there is limited diversity for small benchmarks (P0-P3),
especially when restricting the solutions to the optimal/best-found
ones (0\% optimality gap).
Increasing the optimality gap to 10\% enables \ac{\toolname} to
generate a larger number of program variants.
For both cases, the presence of \ac{PSC}-mitigating constraints
reduces the number of available variants.
The opposite occurs for P3, where \ac{\toolname} generates more
variants compared to \ac{PSC}-unaware code diversification.
This is due to the introduction of additional transformations (random
variable copies) in \ac{PSC}-aware compilation, which increases the
search space and diversification ability of \ac{\toolname}.

For larger benchmarks (P4-P6), \ac{\toolname} generates all the
requested variants (200).
Looking at the diversification time of these benchmarks, we notice a
clear overhead of \ac{PSC}-aware compared to \ac{PSC}-unaware
diversification.
The overhead is up to a slowdown of 55 times for P6 in MIPS32.
For the largest benchmarks, P7-P10, \ac{\toolname} reaches the time
limit (TO) and the number of generated variants is significantly less
than for the \ac{PSC}-insecure variant generation.
Interestingly, increasing the optimality gap to 10\% decreases the
number of generated variants.
As we see in small benchmarks, increasing the optimality gap allows
for non-optimal (according to the model) solutions, which increases
the available variants.
However, increasing the optimality gap, increases also the search
space, which increases the solver overhead for locating solutions.
This results in a reduction of the generated solutions.

We observe similar trends for both MIPS32 and ARM Thumb.
The main difference is that among small benchmarks, only P0 with 0\%
optimality gap appears to lead to reduced diversity in MIPS32.
At the same time, the difference in diversity between secure and
non-secure variants is smaller in MIPS32 (17 compared to 18 in P0) than for
ARM Thumb (1 compared to 2 in P0).
The reason for this is that MIPS32 provides a larger number of
general-purpose registers that may replace vulnerable register
combinations for \ac{ROT} leakages.
%


The lower part of Table~\ref{tab:num_vars} shows the results for
\ac{TSC}-aware diversification.
Here, \ac{\toolname} is able to generate 200 function variants for all
benchmarks.
Interestingly, for C0, the number of variants for \ac{TSC}-unaware
diversification is less than 200 because our \ac{CR} mitigation
introduces performance overhead (see Section~\ref{ssec:rq1}) and thus,
increased diversification capacity.
The diversification-time overhead is less than for \ac{PSC}-aware
diversification, reaching up to a slowdown of eight times (C0, 0\%
optimality gap, MIPS32).
In all cases, \ac{\toolname} was able to generate 200 variants in less
than 30 seconds.



To summarize, we observe a clear effect on the diversification time
and available diversity in \ac{\toolname} compared to \ac{SCA}-unaware
code diversification.
This effect is more significant in \ac{PSC}-aware diversification,
where there is a general decrease in diversity and increase in the
diversification-time slowdown.
\ac{TSC}-aware diversification appears to affect mainly
diversification time, whereas in some cases, the \ac{CR}
countermeasure increases the available diversity.
Nonetheless, in almost all cases, \ac{\toolname} generates program
variants within ten minutes.


\subsection{Effect of Security Constraints on Code-Reuse Attacks}
\label{ssec:rq4}

\begin{table*}[h]
  \centering
  \caption{\label{tab:gadgets} \ac{CRA} gadget-overlap rate in pairs
    of variants; \secured denotes secure variants and \insecured
    non-secure variants}
  \setlength{\tabcolsep}{3.5pt}
  \begin{tabular}{|l|rrr|rrr|rrr|rrr|rrr|rrr|rrr|rrr|}    
    \hline
    \multirow{4}{*}{Prg} & \multicolumn{12}{c|}{ARM Cortex M0} & \multicolumn{12}{c|}{Mips32} \\\cline{2-25}
        & \multicolumn{6}{c|}{0\%} & \multicolumn{6}{c|}{10\%} & \multicolumn{6}{c|}{0\%} & \multicolumn{6}{c|}{10\%} \\\cline{2-25}
    & \multicolumn{3}{c|}{\secured}& \multicolumn{3}{c|}{\insecured}& \multicolumn{3}{c|}{\secured} & \multicolumn{3}{c|}{\insecured}& \multicolumn{3}{c|}{\secured} & \multicolumn{3}{c|}{\insecured}& \multicolumn{3}{c|}{\secured} & \multicolumn{3}{c|}{\insecured}\\\cline{2-25}
    &	0&20&100	&	0&20&100	&	0&20&100	&	0&20&100	&	0&20&100	&	0&20&100	&	0&20&100	&	0&20&100\\\hline
P0	&	-&-&-	&	-&-&\textbf{100}	&	23&-&\textbf{77}	&	21&-&\textbf{79}	&	\textbf{100}&-&-	&	\textbf{100}&-&-	&	\textbf{100}&-&-	&	\textbf{100}&-&-\\
P1	&	-&\textbf{53}&47	&	-&\textbf{78}&23	&	14&\textbf{55}&31	&	12&\textbf{71}&16	&	\textbf{89}&-&11	&\textbf{94}&-&6	&	\textbf{95}&5&-	&	\textbf{94}&6&-\\
P2	&	-&\textbf{100}&-	&	-&\textbf{100}&-	&	3&\textbf{66}&31	&	5&\textbf{64}&32	&	\textbf{93}&6&1	&	\textbf{94}&6&1	&	\textbf{97}&3&-	&	\textbf{97}&2&-\\
P3	&	-&\textbf{71}&29	&	-&\textbf{73}&27	&	2&\textbf{83}&15	&	11&\textbf{81}&9	&	\textbf{99}&1&-	&	\textbf{99}&1&-	&	\textbf{99}&1&-	&	\textbf{99}&1&-\\
P4	&	-&\textbf{86}&14	&	-&\textbf{75}&25	&	5&\textbf{87}&7	&	10&\textbf{83}&7	&	\textbf{100}&-&-	&	\textbf{100}&-&-	&	\textbf{99}&-&-	&	\textbf{99}&1&-\\
P5	&	-&\textbf{86}&14	&	-&\textbf{75}&25	&	5&\textbf{87}&7	&	10&\textbf{83}&7	&	\textbf{100}&-&-	&	\textbf{100}&-&-	&	\textbf{99}&-&-	&	\textbf{99}&1&-\\
P6	&	-&\textbf{90}&10	&	-&\textbf{87}&13	&	8&\textbf{86}&6	&	15&\textbf{80}&4	&	\textbf{100}&-&-	&	\textbf{100}&-&-	&	\textbf{99}&1&-	&	\textbf{99}&1&-\\
P7	&	1&\textbf{92}&8	&	-&\textbf{95}&5	&	6&\textbf{87}&7	&	7&\textbf{88}&5	&	\textbf{98}&2&-	&	\textbf{100}&-&-	&	\textbf{99}&1&-	&	\textbf{100}&-&-\\
P8	&	2&\textbf{88}&11	&	-&\textbf{93}&6	&	19&\textbf{75}&6	&	3&\textbf{91}&5	&	\textbf{97}&2&1	&	\textbf{100}&-&-	&	\textbf{98}&2&1	&	\textbf{100}&-&-\\
P9	&	-&\textbf{83}&17	&	32&\textbf{63}&5	&	10&\textbf{85}&5	&	46&\textbf{50}&4	&	\textbf{78}&10&12	&\textbf{100}&-&-	&	\textbf{95}&1&4	&	\textbf{99}&1&-\\
P10	&	\textbf{57}&42&2	&	-&\textbf{95}&5	&	\textbf{75}&22&2	&	\textbf{64}&35&1	&	\textbf{79}&19&1	&	\textbf{94}&6&-	&	\textbf{66}&26&8	&	\textbf{99}&1&-\\
\hline
C0	&	36&\textbf{61}&3	&	-&45&\textbf{55}	&	43&\textbf{54}&3	&	\textbf{55}&30&15	&	29&\textbf{68}&2	&\textbf{95}&-&5	&	\textbf{80}&18&2	&	\textbf{86}&14&-\\
C1	&	34&\textbf{66}&-	&	-&\textbf{100}&-	&	\textbf{95}&4&1	&	\textbf{93}&6&1	&	29&\textbf{71}&-	&	41&\textbf{59}&-	&	\textbf{92}&5&2	&	\textbf{95}&3&2\\
C2	&	-&\textbf{69}&31	&	-&\textbf{68}&32	&	\textbf{42}&37&21	&	\textbf{53}&33&13	&	\textbf{90}&10&-	&\textbf{82}&18&-	&	\textbf{92}&8&-	&	\textbf{80}&20&-\\
C3	&	18&\textbf{56}&26	&	-&1&\textbf{99}	&	\textbf{93}&6&2	&	\textbf{93}&4&3	&	8&\textbf{92}&-	&	-&\textbf{100}&-	&\textbf{97}&1&2	&	\textbf{96}&3&2\\
C4	&	\textbf{57}&41&1	&	-&\textbf{97}&3	&	\textbf{96}&4&-	&	\textbf{95}&5&-	&	\textbf{94}&6&-	&	\textbf{94}&6&-	&	\textbf{100}&-&-	&	\textbf{99}&1&-\\
\hline
  \end{tabular}
\end{table*}

This section evaluates the effect of \ac{SCA}-aware diversification on
the effectiveness against \acp{CRA}.
To evaluate the effectiveness of \ac{SecDivCon} against
\acp{CRA}, we measure the rate of code-reuse gadgets that are
relocated or transformed among different variants.
We perform this evaluation at the generated binary ELF~\cite{elf}
files.
This evaluation uses ROPgadget\footnote{ROPgadget:
  \url{https://github.com/JonathanSalwan/ROPgadget}}, a tool that
extracts code-reuse gadgets from a binary and Capstone, a
lightweight disassembly framework.
We extract the gadgets from the \texttt{.text} section of the
generated ELF files.
Similarly to previous
work~\cite{homescu_profile-guided_2013,pappas_smashing_2012},
we assess the gadget-overlap rate $srate(p_i,p_j)$ for each pair of
variants $p_i,p_j \in S$ in the set of generated variants, $S$, to
evaluate the effectiveness of \ac{\toolname} against \acp{CRA}.
This metric returns the rate of the gadgets of variant $p_i$ that
appear at the same address in the second variant $p_j$.
The procedure for computing $srate(p_i,p_j)$ is as follows: 1) run
ROPgadget on variant $p_i$ to find the set of gadgets $gad(p_i)$ in
variant $p_i$, and 2) for every $g\in gad(p_i)$, check whether there
exists a gadget identical to $g$ at the same address in the second
variant $p_j$.
Before the comparison, we remove all \ac{NOP} instructions.
%
%
The smaller the $srate$ is, the fewer gadgets are shared among program
variants, and thus, the highest the effect against \acp{CRA}.
Note that $srate$ does not check the semantic equivalence of the
gadgets, and hence, there may be false negatives, namely pairs of
gadgets that are syntactically different but semantically equivalent.
We use a time limit of ten minutes and an upper bound on the number of
variants to generate because of the increasing complexity of the
pairwise gadget-overlap rate that depends on all pairs of generated
variants.

Table~\ref{tab:gadgets} shows the rate of shared code-reuse gadgets
among the generated variants for ARM Cortex M0 and MIPS32.
For each processor, Table~\ref{tab:gadgets}, shows the results for
two configurations that allow variants to introduce at most 0\% to 10\%
execution-time overhead.
We compare \ac{SCA}-aware variants (\secured) and \ac{SCA}-unaware
variants (\insecured).
For each of these cases, Table~\ref{tab:gadgets} shows the $srate$,
i.e.\ rate of pairs of variants, in the form of a histogram with three
buckets.
The buckets represent the rate of variant pairs that share 1) $0\%$ of
their gadgets (0 in Table~\ref{tab:gadgets}), 2) $(0\%, 20\%]$ of the
  gadgets (20 in Table~\ref{tab:gadgets}), or 3) $(20\%, 100]$ of the
    gadgets (100 in Table~\ref{tab:gadgets}).
The goal of \ac{\toolname} is to generate variants that share as few
gadgets as possible, i.e. the variant pairs share no gadgets (0 in
Table~\ref{tab:gadgets}).

In Table~\ref{tab:gadgets}, we observe a general difference between
the two processors, with \ac{\toolname} achieving lower gadget
survival rate for MIPS32 than ARM Cortex M0.
We describe the results for the two processors in the following.

In ARM Cortex M0, with 0\% allowed execution-time overhead, for both
\ac{SCA}-aware and \ac{SCA}-unaware diversification, the mode of the
pairwise survival rate for the majority of the benchmarks lies within
$(0\%, 20\%]$.
For \ac{SCA}-aware diversification for 13 benchmarks the mode of the
distribution is under $(0\%, 20\%]$ and for two is 0\% (P10 and C4).
The results for \ac{SCA}-unaware diversification are similar, with P9
having improved gadget elimination and P10, C0, C3, and C4 having
reduced gadget-elimination ability than \ac{\toolname}.
Increasing the optimality gap to 10\% results in reduced survival rate
(improvement).
In particular, for \ac{SCA}-aware diversification, five benchmarks
have a distribution with the mode in 0\%, ten have their mode
under $(0\%, 20\%]$, and one under $(20\%, 100\%]$.
Here, the results for \ac{\toolname} are similar to \ac{SCA}-unaware
diversification, with C0 showing better results in  \ac{SCA}-unaware
diversification.

In contrast, for MIPS32, most experiments (apart for C0, C1, and C3
with 0\% optimality gap) have their mode under 0\% survival rate,
which means that the majority of variant pairs do not share any
gadgets.
The reason why MIPS32 appears to achieve higher gadget
relocation/diversification is the characteristics of the architecture
with many general purpose registers.
ARM Cortex M0, on the other hand, has significantly fewer
general-purpose hardware registers and multiple 2-address instructions
that are highly constrained.

To summarize, the results show relatively low gadget survival rate for
both ARM Cortex M0 and MIPS32, whereas, this survival rate does not
appear to increase (worse) for \ac{SCA}-aware diversification.
This means that combining \ac{SCA} mitigations with diversification
against \acp{CRA} does not reduce the mitigation capability of
fine-grained diversification against \acp{CRA}.

\section{Discussion}

This section discusses the application of \ac{\toolname} against more
advanced attacks and the potential extension of \ac{\toolname} to
support additional mitigations.

\paragraph{Whole-Program Mitigation:}

  Our threat model considers static gadget-based code-reuse attacks,
  such as \ac{ROP} attacks~\cite{shacham_geometry_2007}.
  \ac{\toolname} proposes a fine-grained function-level
  diversification approach as a mitigation against these attacks.
  Combining the generated variants for each function allows for
  whole-program diversification~\cite{tsoupidi2021constraint}.
  \ac{RILC}~\cite{tran_expressiveness_2011} attacks where the gadgets
  correspond to entire functions may be defeated by combining
  whole-program diversification with function shuffling and/or
  coarse-grained diversification approaches, such as \ac{ASLR}.

  \paragraph{Advanced Attacks:}
  Advanced code-reuse attacks, such as
  \ac{BROP}~\cite{bittau_hacking_2014}, may use a memory vulnerability
  to read the program memory and find gadgets dynamically in the
  diversified code.
  \ac{BROP} attacks read the program memory using a memory
  vulnerability and depend on the reset of the system after a system
  crash.
  %
  An efficient approach against \ac{BROP} is
  re-randomization~\cite{harm} that may be performed at boot
  time~\cite{caballero_avrand_2016}.
  Runtime re-randomization switches program variants at runtime at an
  interval within which the attacker should not be able to complete an
  attack.
  The main drawbacks of re-randomization is that 1) it may lead to
  high memory footprint for the
  binary~\cite{cabrera_arteaga_multi-variant_2022}, which may be
  forbidding in resource-constrained devices, and 2) it contributes to
  additional performance overhead.
  Nonetheless, \ac{\toolname} performs fine-grained automatic
  diversification that may be used in a re-randomization scheme,
  enabling improved protection against advanced code-reuse attacks.

  Apart from classical power analyses, such as \ac{DPA} and \ac{CPA},
  recently, the advancement of deep learning has allowed more powerful
  attacks.
  \citet{ngo_breaking_2021} show that advanced randomization
  techniques, such as plaintext shuffling, are
  vulnerable~\cite{ngo_breaking_2021,ngo_side-channel_2021-1}, when
  the implementation leaks secret values.
  They also show that first-order masking can be defeated with
  deep-learning based analysis, however, the masking property is
  preserved at the source-code level, thus \ac{ROT} or \ac{MRE}
  leakages may be present after
  compilation~\cite{athanasiou_automatic_2020-1}.
  We leave the evaluation of our approach against these attacks as 
  future work.
  
  \paragraph{Implement Additional Mitigations:}
  \ac{\toolname} combines code diversification and side-channel
  attack mitigations to protect embedded devices.
  However, additional mitigations may be necessary to protect a
  device against other types of attacks.
  An essential step for combining different mitigations is to
  determine whether these mitigations are conflicting.
  In case they are, the designer may describe the new mitigations as
  constraints to extend the constraint model of \ac{\toolname}.
  This allows \ac{\toolname} to generate secure code. 
  %
  \paragraph{Security Policy:}

  \ac{\toolname} is suitable for analyzing functions that process
  secret values and hence, are vulnerable to side-channel attacks.
  The security policy defines, which of the input values are
  \texttt{secret}, and which of them are \texttt{public} or
  \texttt{random}.
  %
  %
  The type of each value depends on the usage of the respective
  algorithm in a security-critical application.
  \ac{\toolname} requires only annotating the function arguments and
  applies a type-inference algorithm to extract the type of each
  intermediate variable.

\section{Related Work}
\label{sec:relwork}

\begin{table}
\caption{\label{tbl:relwork} Related work; CRA stands for code-reuse
  attacks; TSC stands for timing side-channel attacks; MS stands for
  memory safety; PSC stands for power side-channel attacks; IL stands
  for interrupt-latency SCA; Div stands for diversification; Obf
  stands for obfuscation; CFI stands for control-flow integrity; CT
  stands for constant-time discipline; SM stands for software masking;
  BB stands for basic-block balance; RR stands for re-randomization;
  HWCFI stands for hardware-assisted CFI; PO corresponds to the upper
  bound of the performance overhead; SDC stands for \ac{\toolname}. }
\centering
  \begin{tabular}{|l|x{1.5cm} c c x{1.4cm}|}
    \hline
    Pub.                  & Attack   & Mitigation  & PO    & Target\\
    \hline
    \cite{homescu_profile-guided_2013} & CRA & Div        & 25\%  & x86 \\
    \cite{pappas_smashing_2012}  &   CRA    & Div         & 0\%   & x86 \\
    \cite{crane_thwarting_2015}  &   TSC    & Div         & 8x    & x86  \\
    \cite{rane_raccoon_2015}     &   TSC    & Obf         & 16x   & x86 \\
    \cite{pastrana_avrand_2016}  &   CRA    & Div, RR     & -     & AVR \\
    \cite{abera_c-flat_2016-1}   &   CRA    & CFI         & $\sim$80\% & ARM \\
    \cite{nyman_cfi_2017}        &   CRA    & CFI         & 5x    & ARM \\
    \cite{zinzindohoue_hacl_2017-1} &TSC, MS& CT          & -     & C   \\
    \cite{koo_compiler-assisted_2018} & CRA & Div, RR     & 7\%   & x86 \\
    \cite{salehi_microguard_2019-1} & CRA   & Div,CFI     & 70\%  & ARM \\
    \cite{shelton_rosita_2021-1} &   PSC    & SM          & 64\%  & ARM  \\
    \cite{winderix_compiler-assisted_2021}& TSC, IL& BB   & 60\%  & MSP430\\
    \cite{harm}                  &   CRA    & Div, RR     & 6\%   & ARM  \\
    \cite{borrello_constantine_2021}&TSC    & CT          & 5x    & x86  \\
    \cite{fu_fh-cfi_2022}        &   CRA    & HWCFI       & 24\%  & ARM  \\
    \hline
  \rowcolor{white!90!black}
    SDC                        &  TSC, PSC, CRA & Div, SM/BB & 70\%$^{\mathrm{a}}$ &  MIPS, ARM\\
    \hline
    \multicolumn{5}{p{0.45\textwidth}}{$^{\mathrm{a}}$Diversification overhead is controlled}
  \end{tabular}
\end{table}

This section presents the related work with regards to \aclp{CRA} and
\aclp{SCA}.
Table~\ref{tbl:relwork} shows a representative subset of
compiler-based or binary-rewrite contributions against \acp{CRA} and
\acp{SCA} in the literature.
For each of these works, Table~\ref{tbl:relwork} shows the publication
citation reference (Pub.), the attack it is mitigating (Attack), the
type of mitigation the publication is proposing (Mitigation), the
maximum performance overhead the approach introduces (PO), and the
target language/architecture (Target).

%

  \subsection{Mitigations against Code-Reuse Attacks}

In the literature, there are two main approaches against \acp{CRA},
software diversification and \ac{CFI}.

Automatic software diversity has been proposed as an efficient
mitigation against \acp{CRA}~\cite{larsen_sok_2014}.
Many software diversification approaches target x86
systems~\cite{homescu_profile-guided_2013,pappas_smashing_2012,koo_compiler-assisted_2018},
while others target embedded
systems~\cite{pastrana_avrand_2016,salehi_microguard_2019-1,tsoupidi2021constraint,harm}.
%
%
The main characteristic of these approaches is that they lead to
relatively low performance overhead.
For example, fine-grained diversification approaches may lead to 0\%
performance
overhead~\cite{pappas_smashing_2012,tsoupidi2021constraint}.

Re-randomization
approaches~\cite{harm,pastrana_avrand_2016,koo_compiler-assisted_2018}
repeat the randomization process in specific timing intervals to
protect against advanced \acp{CRA}, such as
JIT-ROP~\cite{snow_just--time_2013}, \ac{BROP}, and side-channel-based
diversification deciphering~\cite{seibert_information_2014}.
These approaches may introduce additional binary-size
overhead~\cite{cabrera_arteaga_multi-variant_2022} and performance
overhead.
However, this performance overhead is typically low, for example,
HARM~\cite{harm} introduces up to 6\% additional overhead.

\ac{CFI} mitigates \acp{CRA} by ensuring that the dynamic execution of
the program adheres to the intended program control
flow~\cite{burow_control-flow_2017-1}.
Software-based CFI
systems~\cite{abera_c-flat_2016-1,nyman_cfi_2017,salehi_microguard_2019-1}
typically result in high overhead, whereas hardware-assisted methods
may lead to reduced overhead~\cite{fu_fh-cfi_2022}.
However, hardware-assisted \ac{CFI} approaches often depend on
specialized hardware mechanisms~\cite{burow_control-flow_2017-1}.

To summarize, there are multiple approaches to mitigate \acp{CRA}, but
none of them considers or evaluates the effect on mitigations against
\acp{SCA}.
Comparing code diversification and \ac{CFI} approaches, the former
typically lead to lower overhead.
This is the main motivation for selecting code diversification as a
mitigation against \acp{CRA}.

\subsection{Code Hardening Against Side-Channel Attacks}

Software masking is a software approach to mitigate \acp{PSC}.
However, a compiler that translates a program to machine code may
introduce power
leaks~\cite{wang_mitigating_2019-1,shelton_rosita_2021-1,papagiannopoulos_mind_2017,
  athanasiou_automatic_2020-1}.
\citet{wang_mitigating_2019-1} identify leaks in masked implementation
using a type-inference algorithm, and then, perform
register-allocation transformations to mitigate these leaks in LLVM.
Rosita~\cite{shelton_rosita_2021-1} performs an iterative process to
identify power leakages in software implementations for ARM Cortex M0,
with a performance overhead of up to 64\%.
Our recent approach~\cite{tsoupidi2021seccon} based on type
inference~\cite{gao_verifying_2019-1} presents an approach with
execution overhead up to 13\%.
\ac{\toolname} adapts this approach to generate diverse code variants
that preserve software masking.

The constant-time programming discipline~\cite{MolnarPSW05} is a
widely-used programming discipline that prevents \ac{TSC} attacks.
It prohibits the use of secret values in branch decisions, memory
indexes, and variable-latency instructions (such as division in many
architectures).
\citet{borrello_constantine_2021} linearize code to translate a
program to a constant-time equivalent including branches, loops, and
memory accesses.
%
The main drawback of this approach is the introduction of
execution-time overhead of up to five times.
The constant-time programming discipline leads to secure code as it
ensures that there are no secret-dependent timing variations, however
it is restrictive because it does not allow secret-dependent branches
and makes the code difficult to read and
implement~\cite{ngo_verifying_2017-1}.
\citet{barthe_secure_2021} present \ac{CR} programming, an
alternative, more relaxed form of constant-time programming that
allows branches on secret values as long as the diverse execution
paths take identical time to execute.
%
Similarly, \citet{brown_semi-automatic_2022} perform transformations
to balance secret-dependent branches by balancing the branch bodies at
the C level.
\citet{winderix_compiler-assisted_2021} balance secret-dependent
branches with equivalent-latency \acp{NOP} to mitigate \ac{TSC} and Interrupt
Latency Side-Channel Attacks.
The latter attacks distinguish which path of a branch the program
follows based on the latencies of the instructions in each block.
A different approach against timing attacks is
Raccoon~\cite{rane_raccoon_2015}, which uses control-flow obfuscation
to mitigate \ac{TSC} attacks.
However, \citet{joshi2015trading} has shown that obfuscation may
introduce code-reuse gadgets.
Hence, Raccoon may increase the attack surface of \acp{CRA}.
Moreover, this mitigation introduces an overhead of up to 16 times,
which is prohibiting for resource-constraint devices.
\citet{crane_thwarting_2015} present a compiler-based diversification
approach that inserts timing noise to obfuscate cache-based timing
attacks on cryptographic algorithms.
However, this approach introduces a performance overhead of up to 8x,
which is higher than \ac{\toolname} that introduces an overhead of up
to 70\% for generating constant-resource programs.

Finally, HACL* by \citet{zinzindohoue_hacl_2017-1} is a verified cryptographic library that generates C code that is memory safe and constant time.
Although memory safety hinders memory corruption vulnerabilities in
the generated library, HACL* does not prohibit memory vulnerabilities
in the rest of the code, which may enable \acp{CRA}.
Thus, mitigations against \acp{CRA} may still be necessary.

In summary, there are compiler-based and binary rewriting approaches to mitigate \ac{PSC} attacks and \ac{TSC} attacks; however, none of these approaches are effective against \acp{CRA} and/or consider the effect on \acp{CRA}. 

\section{Conclusion and Future Work}

This paper presents \ac{\toolname} -- a constraint-based approach that
combines code diversification with side-channel mitigations.
It enables the secure-by-design generation of optimized code for
small, predictable hardware architectures.
Our evaluation shows that the introduction of \ac{SCA} mitigation-preserving constraints impacts the scalability of diversification, but it does not have a negative effect against code-reuse attacks.

In future work, we plan to investigate how to improve
\ac{\toolname}'s scalability and extend the \ac{CR}-preserving model
with additional transformations that allow the analysis of
secret-dependent branches that contain bounded loops.

\section*{Acknowledgment}
We would like to thank Jingbo Wang for their support with their tool.
We would also like to thank Andreas Lindner for his support with
verifying \ac{\toolname} using \ac{WCET} analysis for ARM Cortex M0.
In addition, we would like to thank Roberto Casta\~neda Lozano for
his technical support on Unison, \ac{\toolname}'s underlying
constraint-based compiler backend.
Finally, we would like to thank Nicolas Harrand, Amir M. Ahmadian, and
Javier Cabrera Arteaga for their feedback on this paper.
P. Papadimitratos acknowledges the support of the Swedish Science
Foundation (VR) and the Knut and Alice Wallenberg (KAW) Foundation
that funded parts of his work in this context.
E. Troubitsyna acknowledges the support of the Swedish Foundation 
for Strategic Research (SSF) in this work.



\printcredits

\bibliographystyle{cas-model2-names}

\bibliography{bibliography}

\appendix

\section{Path-Finding Algorithm}
\label{app:algorithm}

\begin{figure}[t]
\begin{lstlisting}[style=algomine, mathescape=true] 
GET_PATHS(n, BCFG):
  t.empty() # Queue - First path
  P.empty() # Priority queue - Paths
  t.insert(n) 
  P.insert(t)
  W.empty() # final paths
  while ($\neg$P.isempty() and $\neg$P.hasCycle()):
    p $\leftarrow$ P.top() # Top path
    h $\leftarrow$ p.pop() # Last element of path
    succ $\leftarrow$ BCFG.successors(h)
    if (succ = $\varnothing$): # exit node
      W.push(p)
      P.remove(p)
    elif (succ = {s}):
      p.push(s)
      P.replace(p)
      # if this is a sink, we terminate
      if (W.extend(P).hasSink()):
        return W.extend(P)
    elif (succ = {s1,s2}):
      p1 $\leftarrow$ p.copy()
      p2 $\leftarrow$ p.copy()
      p1.push(s1)
      p2.push(s2)
      P.remove(p)
      P.insert(p1)
      P.insert(p2)
   return W
\end{lstlisting}
\caption{\label{lst:paths}Path extraction} 
\end{figure}

When the branch condition has type \texttt{secret}, i.e.\ depends on a
\texttt{secret} value, \ac{\toolname} performs an analysis to discover
all paths starting from the branch condition (source) to a common node
(sink).
We assume that the program is split into basic blocks, pieces of code
with at most one branch (apart from function calls) at the end of the
block.
To identify all possible paths, we generate the \ac{CFG} between the
basic blocks of the program.

Figure~\ref{lst:paths} shows the algorithm for extracting the paths
that start from a basic block $n$ (the secret-dependent branch), given
the \ac{CFG} (\texttt{BCFG}).
We use two data structures, a priority queue, \texttt{P}, which
contains all paths under analysis, and a queue, \texttt{t} that
represents the current path and starts with the first basic block, $n$
(line 4).
The priority queue uses the block order as the priority, with smaller
numbers having priority. 
At line 5, \texttt{P} is initialized with \texttt{t}.
We store the final results in \texttt{W} (line 6).
At line 7, we start a loop that terminates when there are no paths
left to analyze in \texttt{P} or when we find a cycle.
At lines 8 and 9, we get the top element of the top path from
\texttt{P}.
Subsequently, the algorithm finds all successor nodes in the \ac{CFG},
which correspond to possible basic blocks that follow the current
basic block (line 10).
Then, the algorithm performs different actions depending on the
successor nodes.
First, if the current node, \texttt{h} does not have any successors,
it means that \texttt{h} is an exit node, thus, \texttt{h} is the last
node in the current path.
Lines 12 and 13 add the path to \texttt{W} and remove it from the
paths under analysis.
If \texttt{h} has one successor, \texttt{s}, then we push the
successor to the path and update \texttt{P} (lines 14-16).
Here, we need to check if the new node leads to the current paths
having a sink, i.e. the same final node (line 17).
The last case is when the branch is conditional and there are two
possible destinations.
Here, we need to generate two paths \texttt{p1} and \texttt{p2} for
each of the two destinations and insert them to \texttt{P} for further
analysis (lines 20-27).
When the analysis finishes and the algorithm exits the loop, then it
returns \texttt{W}.

This analysis does not support secret-dependent loops
  or public-dependent loops nested in secret-dependent branches.
  The reason for the latter is that \ac{\toolname} does not
  implement a loop-insertion transformation.
  In the presense of these patterns, the compiler will fail to
  generate a balanced program because the constraints are infeasible.
  To handle secret-dependent loops, \ac{\toolname} needs to consider
  additional transformations, such as loop insertion, and converting
  secret-dependent loops to non-terminating
  loops~\cite{soares_side-channel_2023}.
  This is part of future work.





\end{document}